\documentclass[fleqn,usenatbib]{mnras}

\usepackage{newtxtext,newtxmath}
\usepackage[T1]{fontenc}
\usepackage{ae,aecompl}
\usepackage[dvipsnames]{xcolor}
\usepackage[toc,page]{appendix}

\usepackage{graphicx}	% Including figure files
\usepackage{amsmath}	% Advanced maths commands
\title[X-Ray Eclipse Mapping of \textit{Swift} J1858.6$-$0814]{X-ray Eclipse Mapping Constrains the Binary Inclination and Mass Ratio of \textit{Swift} J1858.6$-$0814}

\author[A. H. Knight et al.]{
Amy H. Knight,$^{1}$\thanks{E-mail: amy.knight@physics.ox.ac.uk}
Adam Ingram,$^{2}$
and Matthew Middleton $^{3}$
\\
$^{1}$Department of Physics, Astrophysics, University of Oxford, Denys Wilkinson Building, Keble Road, Oxford, OX1 3RH, UK\\
$^{2}$School of Mathematics, Statistics, and Physics, Newcastle University, Newcastle upon Tyne NE1 7RU, UK\\
$^{3}$School of Physics and Astronomy, University of Southampton, Highfield, Southampton, SO17 1BJ, UK
}

\date{Accepted XXX. Received YYY; in original form ZZZ}

\pubyear{2022}

\begin{document}
\label{firstpage}
\pagerange{\pageref{firstpage}--\pageref{lastpage}}
\maketitle

% Abstract of the paper
\begin{abstract}
X-ray eclipse mapping is a promising modelling technique, capable of constraining the mass and/or radius of neutron stars (NSs) or black holes (BHs) in eclipsing binaries and probing any structure surrounding the companion star. In eclipsing systems, the binary inclination, $i$, and mass ratio, $q$ relate via the duration of totality, $t_{\rm e}$. The degeneracy between $i$ and $q$ can then be broken through detailed modelling of the eclipse profile. Here we model the eclipses of the NS low-mass X-ray binary \textit{Swift} J1858.6-0814 utilising archival \textit{NICER} observations taken while the source was in outburst. Analogous to EXO 0748$-$676, we find evidence for irradiation driven ablation of the companion's surface by requiring a layer of stellar material to surround the companion star in our modelling. This material layer extends $\sim 7000-14000$ km from the companion's surface and is likely the cause of the extended, energy-dependent and asymmetric ingress and egress that we observe. Our fits return an inclination of $i \sim 81^{\circ}$ and a mass ratio $q \sim 0.14$. Using Kepler's law to relate the mass and radius of the companion star via the orbital period ($\sim 21.3$ hrs), we subsequently determine the companion to have a low mass in the range $0.183~M_{\odot} \leq M_{\rm{cs}} \leq 0.372~M_{\odot}$ and a large radius in the range $1.02~R_{\odot} \leq R_{\rm{cs}} \leq 1.29~R_{\odot}$. Our results, combined with future radial velocity amplitudes measured from stellar absorption/emission lines, can place precise constraints on the component masses in this system. 
%\cmtak{Word Count:243}
\end{abstract}

% Select between one and six entries from the list of approved keywords. 
% Don't make up new ones.

%%%%%%%%%%%%%%%%%%%%%%%%%%%%%%%%%%%%%%%%%%%%%%%%%%

%%%%%%%%%%%%%%%%% BODY OF PAPER %%%%%%%%%%%%%%%%%%

\section{Introduction}
X-ray binary systems comprising of a neutron star (NS) or a black hole (BH) in orbit with a secondary star provide means to develop binary evolution models \citep{Steiner2010, PPBinary, Postnov2014}, study accretion processes \citep{Done2010, Zhang2013, Ponti2014} and probe the strong gravity regime \citep{Kaaret1997,Dovciak2004, Stevens2017}. While relatively rare, observations of eclipsing X-ray binaries are fundamental when measuring physical properties, particularly of NSs, for which the equation of state (EoS) remains uncertain \citep{Ozel2016}. These observations can also provide a way to probe ablation processes and properties of the companion star's surroundings \citep{Knight2021}. In such systems, periodic X-ray eclipses occur when the X-ray emitting region is occulted by the companion star \citep{Cominsky1984, Parmar1986}. Naturally, eclipses demand a sufficiently high binary inclination. However, determining the exact inclination angle requires knowledge of the physical properties of secondary star, and the binary mass ratio, $q = M_{\rm{cs}}/M_{\rm{ns}}$, where $M_{\rm{cs}}$ and $M_{\rm{ns}}$ are the mass of the companion star and NS respectively, governs the minimum inclination for which eclipses are observable. In eclipsing X-ray binaries, the binary inclination and mass ratio can be disentangled somewhat since they are related via the duration of totality if the companion star is filling its Roche-Lobe \citep{Horne1985}. Thus, by determining the duration of totality and either the mass ratio or the binary inclination, the remaining parameter can be constrained, providing the orbital period is known. 

The binary inclination can be determined independently of the mass ratio and totality duration through several methods. The inclination of material close to the compact object can be determined by comparing the properties of two sides of the jet \citep{Hjellming1981}, or through modelling the effect of Doppler and gravitational shifts on the X-ray spectrum, typically via the iron line \citep{Fabian1989} but also potentially from the thermal disc spectrum if very good data can be obtained \citep{Parker2019}. Further from the compact object, optical and X-ray disc winds can suggest the inclination of the disc from which they are launched \citep{Ponti2012, Higginbottom2018}. The inclination of the binary orbit itself can be determined from ellipsoidal modulations arising from tidal distortion of the companion star \citep{Casares2014}, or through X-ray eclipse mapping which models the shape and duration of the ingress, egress and totality to determine the binary mass ratio and totality duration \citep{Knight2021}. Eclipse mapping can, therefore, self-consistently constrain the mass ratio, $q$, the binary inclination, $i$ and the totality duration $t_e$, and is particularly beneficial when modelling extended or asymmetric eclipse profiles \citep{Knight2021}.

\textit{Swift} J1858.6$-$0814 is a NS low mass X-ray binary (LMXB) known to exhibit X-ray eclipses that are heavily extended and asymmetric \citep{Buisson2021}. This source was originally discovered as an X-ray transient in October 2018 \citep{Krimm2018} with a variable counterpart observed at optical \citep{Vasilopoulos2018, Baglio2019} and radio \citep{VanDenEijnden2020} wavelengths. Initially, \textit{Swift} J1858.6$-$0814 displayed significant X-ray variability, changing by factors of several hundred within a few hundred seconds \citep{Hare2020}. As a result, \textit{Swift} J1858.6-0814 was described as an analogue of the BH sources V4641 Sgr and V404 Cyg, which were observed to show similarly strong variability during their outbursts \citep{Wijnands2000, Revnivtsev2002, Walton2017, Motta2017}. \textit{Swift} J1858.6$-$0814 transitioned from this so-called flaring outburst state (2018 - 2019) to a steady outburst state (2020) with a more persistent X-ray flux. The first half of 2020 saw the steady-state flux decline and the source has been in quiescence since May 2020 ($\sim 58970$ MJD) \citep{Saikia2020, Parikh2020a}.

During the steady-state, Type I X-ray bursts were detected \citep{Buisson2020} thus identifying \textit{Swift} J1858.6-0814 as a NS LMXB, although no pulsations have been detected. The steady-state enabled the discovery of extended and asymmetric eclipses which appear to depend on energy \citep{Buisson2021}. Analysis by \citet{Buisson2021} of all available \textit{NICER} observations uncovered an average ingress duration of $\sim 100$ s and an average egress duration of $\sim 200$ s. Through simultaneous calculation of the totality duration and orbital period, they respectively determine $t_{\rm e} \sim 4100$ s and $P \sim 21.3$ hrs. Additionally, the eclipse duration to orbital period ratio constrains the inclination to, $i > 70^{\circ}$. \citet{Buisson2021} utilise their calculated orbital period to determine the companion's mass as a function of radius, concluding that the companion must be a sub-giant due to the large inferred stellar radius and inconsistency with the main sequence mass-radius relation of \citet{Demircan1991}.

Here we model X-ray eclipse profiles of \textit{Swift} J1858.6$-$0814 in multiple energy bands, using all available archival \textit{NICER} data. Since the eclipses appear energy-dependent, extended and asymmetric \citep{Buisson2021}, we apply our previously published eclipse profile model \citep{Knight2021}, assuming an X-ray point source, thus allowing us to self-consistently constrain the binary inclination, $i$, the mass ratio, $q$ and the totality duration, $t_e$. We note that the extended ingress and egress duration observed will not allow for a NS radius constraint. These features do, however, enable us to probe the structure of the companion star's surroundings and we infer the presence of an absorbing medium that extends several thousands of kilometres from the stellar surface. This medium is likely the cause of the observed extended and asymmetric eclipse profiles. In Section \ref{sx:data}, we detail our data reduction procedure before presenting stacked energy-resolved eclipse profiles and a fit to the time-averaged spectrum. In Section \ref{sx:ecmodel}, we model the energy-resolved eclipse profiles and use our results to derive a posterior probability distribution for the mass ratio, $q$ and binary inclination, $i$. We discuss our results in Section \ref{sx:diss} and conclude in Section \ref{sx:conclude}.

\section{Data Reduction and Analysis}
\label{sx:data}
We consider all available archival \textit{NICER} observations of \textit{Swift} J1858.6$-$0814 during outburst; these are all ObsIDs beginning with 120040, 220040, 320040 or 359201. These observations occurred between November 2018 and July 2020, thus containing detections from both the flaring and steady outburst states.

\subsection{Data Reduction}
The data are reduced using the \textit{NICER} data reduction software \textsc{nicerdas} V008 (HEASoft V6.29, CALDB 20210707), keeping most filtering criteria to their default values (e.g \citealt{Buisson2021}). We include data taken at low Sun angle by following the procedure of  \citet{Buisson2021} and relax the undershoot rate limit to allow up to 400 cts/s (per FPM). At low Sun angles, optical loading is relatively high, which can deteriorate the response and raise the background at energies $\lesssim 0.4$ keV. This does not impact our timing analysis or modelling of eclipses between $0.4 - 10.0$ keV. We again follow \citet{Buisson2021} to carefully remove any achromatic dips arising from occultation of the detector plane by parts of the ISS as there are some instances in which these are not filtered out by the \textit{NICER} pipeline. We barycentre the events and use \texttt{xselect} to extract  $0.4 -10.0$ keV time-averaged spectra and light curves with 1 second time bins, in seven energy bands; $0.4 - 10.0$ keV, $0.4 - 1.0$ keV, $1.0 - 2.0$ keV, $2.0 -4.0$ keV, $4.0 - 6.0$ keV, $6.0 - 8.0$ keV and $8.0 - 10.0$ keV.

\subsection{Eclipse profiles}
\label{sx:EcProf}
\begin{figure}
\centering
\includegraphics[width=\columnwidth]{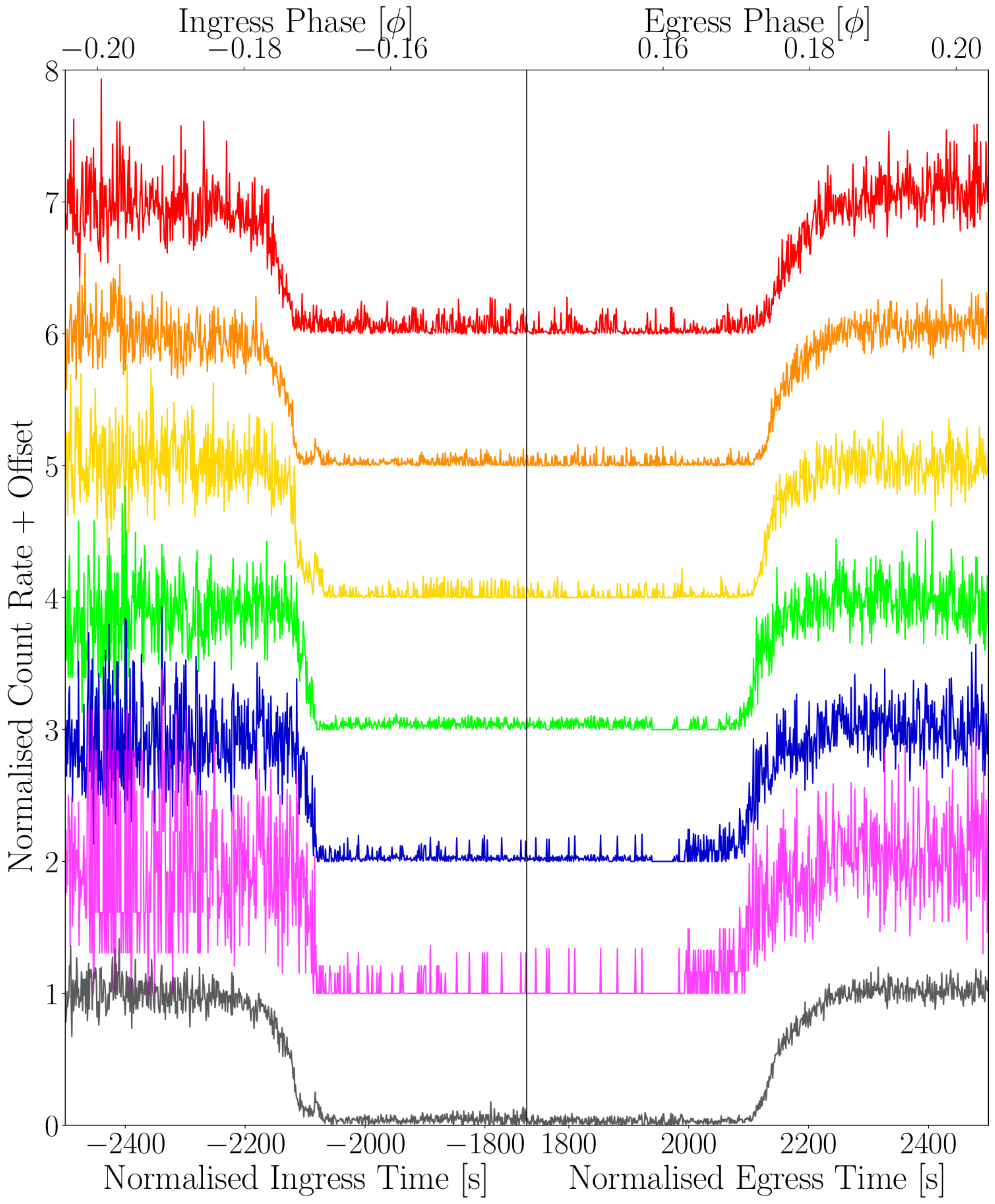}
\vspace{-0.5cm}
\caption{Folded eclipse profiles of \textit{Swift} J1858.6-0814 shown for seven energy bands; $0.4 - 1.0$ keV (red), $1.0 - 2.0$ keV (orange), $2.0 - 4.0$ keV (yellow), $4.0 - 6.0$ keV (green), $6.0 - 8.0$ keV (blue), $8.0 - 10.0$ keV (magenta) and $0.4-10.0$ keV (grey). These have been obtained by folding the extracted light curves on the orbital period $P = 76841.3$ s \citep{Buisson2021}, and dividing through by the mean out-of-eclipse count rate. Note that \textit{NICER} only observed partial eclipses (5 ingresses and 7 egresses). All eclipse profiles are normalised to have a mean out-of-eclipse level of $1.0$ and a mean totality level of $\sim 0.05$ (the totality level is not 0.0 due to a low in-eclipse background count rate). The light curves are shifted such that the time at the centre of the eclipse is at 0.0 seconds ($\phi = 0.0$). The eclipse profiles are shown with a vertical offset. These are $+0.0$ (grey), $+1.0$ (magenta), $+2.0$ (blue), $+3.0$ (green), $+4.0$ (yellow), $+5.0$ (orange) and $+6$ (red).}
\label{fig:ECProfs}
\end{figure}

\begin{figure*}
\centering
\includegraphics[width=\textwidth]{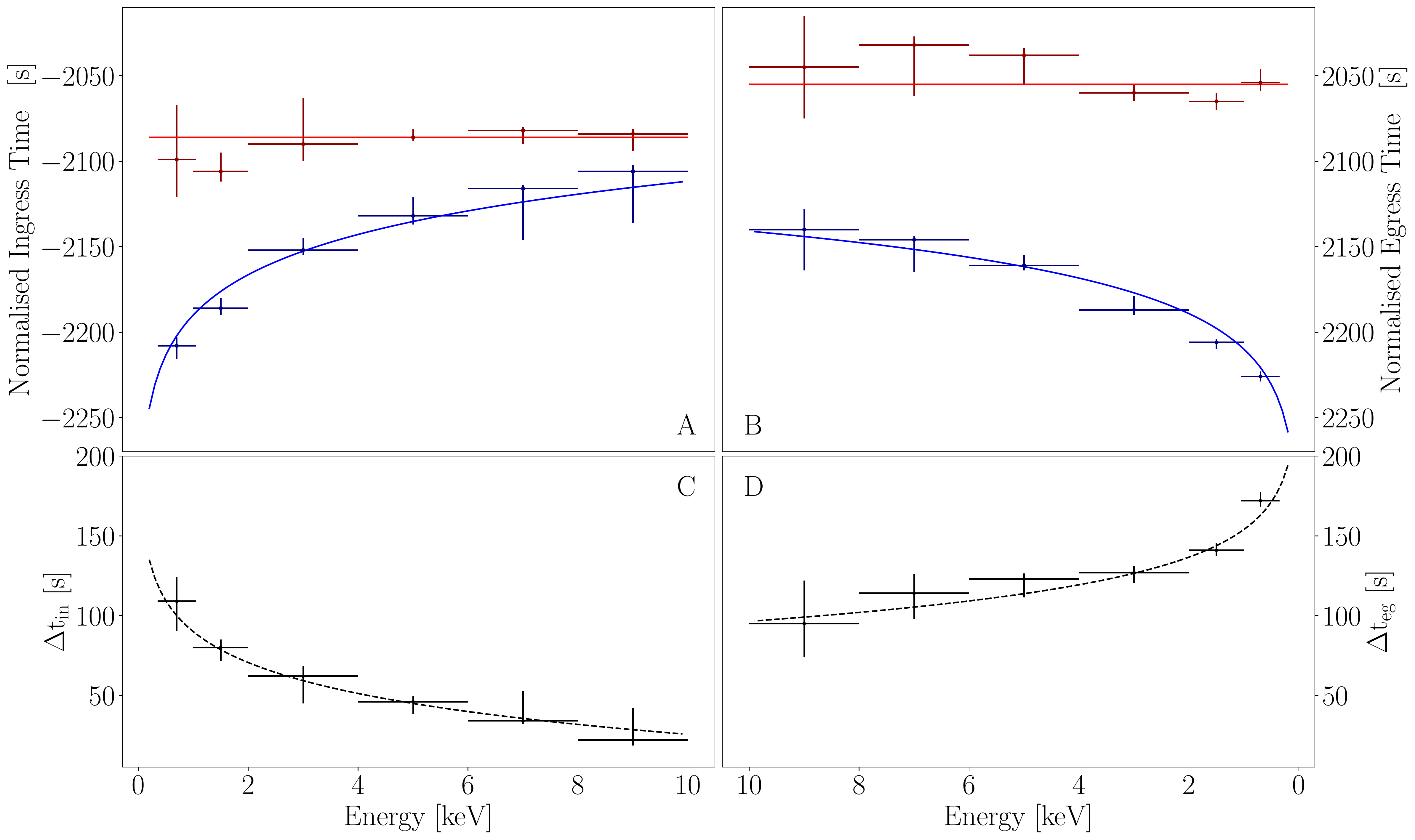}
\vspace{-0.5cm}
\caption{Eclipse transition times, $\rm{t}_{90}$ and $\rm{t}_{10}$, as functions of energy. A time during the eclipse transition, $\rm{t}_{x}$, is defined as the time at which the count rate first passes $x$ per cent of the mean out-of-eclipse level (e.g. \citealt{Knight2021}). Panel A: $\rm{t}_{90}$ (blue) represents the starts of the ingress, which starts earlier for lower energies. The start of totality, $\rm{t}_{10}$ (red), is approximately independent of energy. Panel B: The end of totality (red) is approximately independent of energy, but the end of the egress (blue) depends on energy, ending later for softer X-rays. For the ingress and egress, $\rm{t}_{90}$ and $\rm{t}_{10}$ have been measured from the folded eclipse profiles in Fig. \ref{fig:ECProfs}. Panels C and D: The eclipse transition duration as a function of energy. The duration decreases with increasing photon energy and the egress is consistently longer in duration than the ingress. Note that both sets of axes have been reversed to assist the comparison between the ingress and egress.
%Measured eclipse times, $\rm{t}_{90}$ and $\rm{t}_{10}$, as functions of energy for the ingress and egress; panels A and B respectively. We define the times $\rm{t}_{90}$ and $\rm{t}_{10}$ as, respectively, the time at which the count rate is first at 90 and 10 per cent of its mean out-of-eclipse level. Times are measured from each of the eclipse profiles in Fig. \ref{fig:ECProfs}. Panel A: Ingress start times (blue) increase with energy indicating that eclipses start later for higher photon energies. Ingress end times (red) mark the start of totality and are approximately independent of energy. Panel B: The end of totality that marks the start of the egress is approximately independent of energy (red), but the egress ends later for softer X-rays (blue). Note that both sets of axes have been reversed to aid the comparison of the egress with the ingress. Panels C and D respectively show that the ingress and egress duration decrease with increasing photon energy. We also see that the egress is longer in duration than the ingress.
}
\label{fig:T90}
\end{figure*}

To obtain eclipse profiles in each energy band, we fold the data on the orbital period $P = 76841.3^{+1.3}_{-1.4}$ s ($\approx 21.3$ hours) \citep{Buisson2021} and divide through by the mean out-of-eclipse count rate. Fig. \ref{fig:ECProfs} shows the resulting eclipse profiles for all seven energy bands, displayed with vertical offsets for visual clarity. Note that the eclipse profile in Fig. \ref{fig:ECProfs} arises from partial eclipses only since the eclipse duration of $\sim 4100$ s \citep{Buisson2021} is too long to be observed in full by \textit{NICER}. We find 5 ingresses ($\sim$ orbital cycles 39, 43, 47, 52 and 56) and 7 egresses ($\sim$ orbital cycles 28, 32, 38, 42, 50, 51 and 55), where the zeroth orbital cycle is defined to coincide with the onset of the steady state at $\sim 58885$ MJD \citep{Buisson2021}. Since we divide through by the mean out-of-eclipse count rate, the stacked eclipse profile for each energy band reaches unity during out-of-eclipse phases. The totality, however, is $\sim 0.05$ and not zero due to a background contributed, low in-eclipse count rate. The eclipse profiles are also shifted such that the time at the centre of the eclipse is $0.0$ s (orbital phase $\phi = 0.0$).

\textit{Swift} J1858.6-0814 shows two distinct outburst states - the flaring state and the steady-state divided at MJD 58885 (see in Fig. 1 of \citealt{Buisson2021}). The eclipses are only readily apparent in the light curves of the more recent (2020) steady-state, which show partial eclipses in the form of 5 ingresses and 7 egresses. However, the flaring state observations are also consistent with including eclipses at the orbital phases expected from analysing the steady-state. The extreme flaring and frequent telemetry drop-outs make it difficult to identify eclipses in the light curves during the flaring state, but folding all observations on the orbital period derived from the steady-state observations reveals that the flaring state count rate is always consistent with the background during expected totality phases \citep{Buisson2021}. Despite this, there are no clear ingresses or egresses in the flaring state data since they all happen to occur within telemetry gaps.

We see from Fig. \ref{fig:ECProfs} that the eclipse ingress and egress are both heavily extended in time, with the egress appearing to be longer in duration than the ingress (consistent with the analysis of \citealt{Buisson2021}). We also see the eclipse profile shape, and therefore ingress and egress duration appears to depend on photon energy. To investigate this further, we plot estimates of the start and end times of ingress and egress as a function of energy in Fig. \ref{fig:T90}. Here, following \citealt{Knight2021}, a time during an eclipse transition, $\rm{t}_{x}$, is defined as the time at which the count rate first passes $x$ per cent of the mean out-of-eclipse level and remains past it for $\sim 5$ s. Therefore, when applied to the ingress and egress, $x = 90$ measures the ingress start time and the egress end time, while $x = 10$ measures the totality start and end times.
%Here, following \citet{Knight2021}, we define the times t$_{90}$ and t$_{10}$ as those at which the count rate is first at 90 and 10 per cent of the mean out-of-eclipse level, respectively. Therefore, for the ingress, t$_{90}$ marks the start and t$_{10}$ the end, whereas, for the egress, t$_{10}$ marks the start and t$_{90}$ the end. To account for stochastic variability, we define these times as the time when the average count rate first passes the desired percentage and stays past it for at least five seconds \citep{Knight2021}.

The t$_{90}$ times depend strongly on energy. For the ingress, t$_{90}$ increases with photon energy and the egress t$_{90}$ mirrors this, decreasing with photon energy. This behaviour is well described by a logarithmic equation (solid blue trend line in Figs. \ref{fig:T90}A and \ref{fig:T90}B). For both the ingress and egress, t$_{10}$ is approximately independent of energy, although some variations are observed. These variations likely result from the presence of a background contributed, fluctuating, low in-eclipse count rate. The t$_{10}$ and t$_{90}$ behaviour is equivalent to the duration of both the ingress and egress decreasing with increasing energy. This is shown explicitly in Figs. \ref{fig:T90}C and \ref{fig:T90}D respectively for the ingress and egress. These plots additionally confirm that the egress is longer in duration than the ingress. We measure the ingress duration to be $\rm{t_{10, in, 0.4-10.0 keV} - t_{90, in, 0.4-10.0 keV}} \approx 106$ s and the egress duration to be $\rm{t_{90, eg, 0.4-10.0 keV} - t_{10, eg 0.4-10.0 keV}} \approx 174$ s. 

These extended, asymmetric, energy-dependent eclipse transitions are similar to those observed in EXO 0748$-$676 by \textit{EXOSAT} \citep{Parmar1991}, \textit{RXTE} \citep{Wolff2009} and \textit{XMM-Newton} \citep{Knight2021}, and can be explained by the presence of an ionised layer of material around the companion star \citep{Knight2021}. As our sight-line passes closer to the companion star, the column density of the material layer increases thus causing an energy-dependent drop in flux during the ingress. A sufficiently high column density is, therefore, achieved close to the companion's surface and results in energy independence at the start and end of totality. The eclipse asymmetry can be explained if the absorbing medium trails behind the companion star as a result of the stars orbital motion. Given the remarkable similarity between the eclipses in EXO 0748$-$676 and \textit{Swift} J1858.6-0814, here we consider the same model developed in \citet{Knight2021} for EXO 0748$-$676. Since the material layer is much larger than the NS, we approximate the X-ray source as a point source.

\subsection{Fit to the Time Averaged Spectrum}
\label{sx:SpectralFit}

\begin{figure}
\centering
\includegraphics[width=\columnwidth]{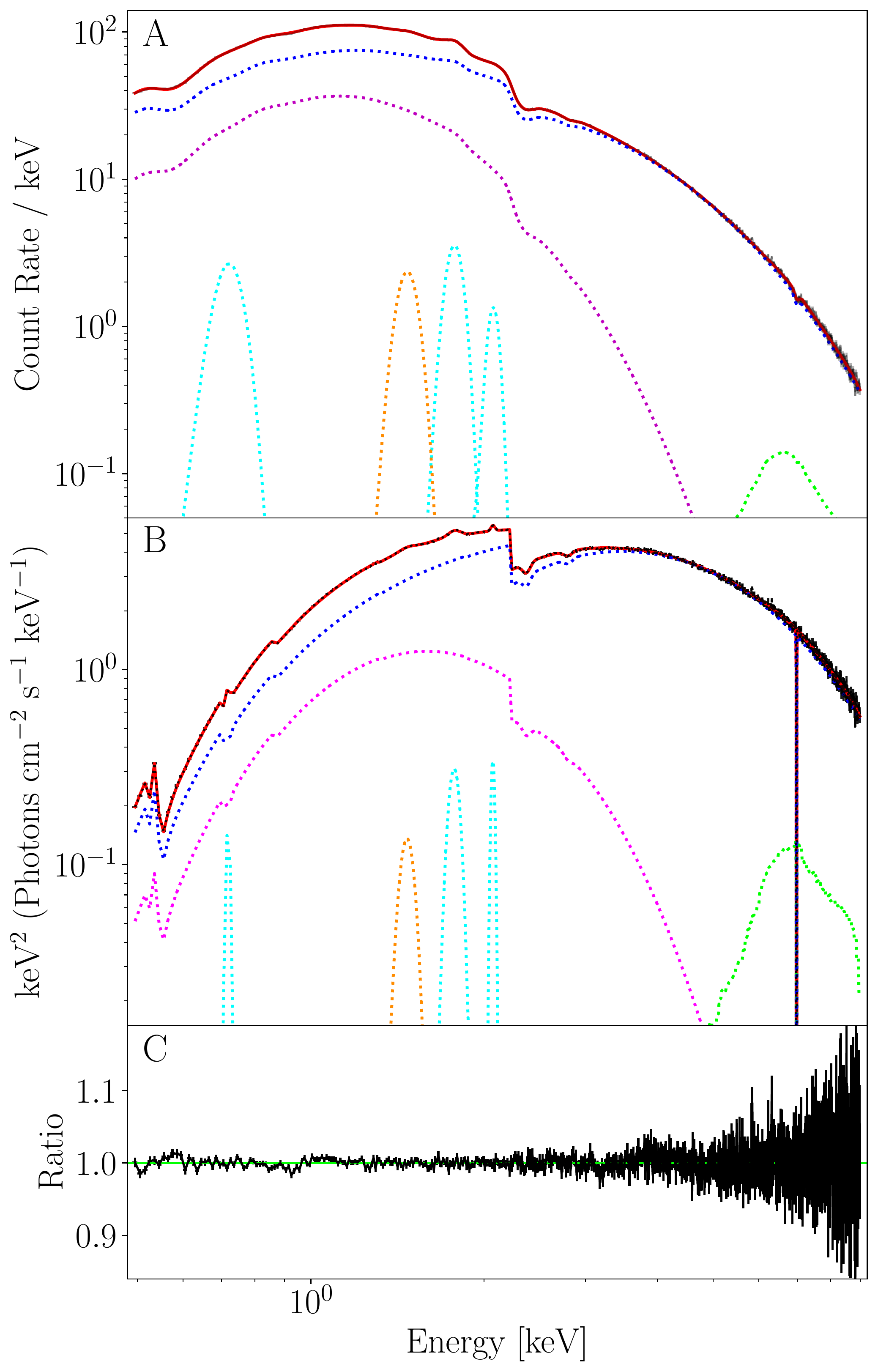}
\vspace{-0.5cm}
\caption{Panels A (best-fitting folded spectrum) and B (best-fitting unfolded spectrum): The time-averaged spectrum of \textit{Swift} J1858.6-0814 (black) fit with the multi-component model (red) detailed in Section \ref{sx:SpectralFit}. The individual components coloured blue and magenta originate from the accretion disc and the NS surface respectively. Also shown are the Laor iron line profile (green), astrophysical emission lines (cyan) and an emission line likely originating from \textit{NICER} calibration systematics (orange). These emission lines correspond to $0.718$ keV (Fe XVIII L$\beta$), $1.47$ keV (Al K$\alpha$, orange) , $1.77$ keV (Si VIII K$\alpha$), and $2.10$ keV (P XIV K$\beta$). Model parameters are summarised in Table \ref{tb:SpecFit}. Panel C: The ratio data/folded model.
%A fit to the time-averaged spectrum of \textit{Swift} J1858.6-0814 (black) using the multi-component model (red) described in Section \ref{sx:SpectralFit}. Also shown are the model components originating from the NS surface (magenta), the accretion disc (blue), the Laor iron line profile (green), astrophysical emission lines (cyan) and an emission line likely originating from \textit{NICER} calibration systematics (orange). These are $0.718$ keV (Fe XVIII L$\beta$), $1.47$ keV (Al K$\alpha$, orange) , $1.77$ keV (Si VIII K$\alpha$), and $2.10$ keV (P XIV K$\beta$). Model parameters are summarised in Table \ref{tb:SpecFit}. Panel A shows the best-fitting folded spectrum, panel B shows the best-fitting unfolded spectrum and panel C shows the ratio: data/folded model.
}
\label{fig:SpecFit}
\end{figure}

Using \textsc{xspec} V12.12.0 \citep{Arnaud1996}, we fit the time-averaged spectrum with the model
\begin{equation}
    \texttt{TBabs*(Laor+diskbb+bbody)*E*A*G}.
\end{equation}
Here, \textsc{diskbb} and \textsc{bbody} respectively describe the multi-temperature spectrum originating from the accretion disc and a blackbody spectrum originating from the NS surface. Absorption by the interstellar medium is accounted for by \textsc{tbabs} and assumes the abundances of \citet{Wilms2000}.
%Here, \textsc{tbabs} accounts for absorption by the interstellar medium (we use the abundances of \citealt{Wilms2000}), and \textsc{diskbb} is a multi-temperature accretion disc spectrum. We model the spectrum from the NS surface as a blackbody (\textsc{bbody}). 
The spectrum requires an emission contribution from Fe K$\alpha$ which is modelled as a \textsc{laor} emission line with $E = 6.57$ keV \citep{Laor1991}. The \textsc{laor} component models the Fe K$\alpha$ line as a relativistically smeared line, assuming a delta function in the rest frame. It would be more precise to instead use a full X-ray reflection model such as \textsc{relxill} \citep{Garcia2014}, which also accounts for effects such as electron scattering and absorption edges. We attempted spectral fits using \textsc{relxill}, \textsc{xillver} and \textsc{xillvercp} (flavours of the \textsc{relxill} model), but found that these models were not well suited to modelling the softer X-ray components of the spectrum. Nonetheless, the \textsc{laor} component captures the asymmetrically broadened shape of the line indicative of relativistic smearing from a highly inclined disc.
% Note that more precise modelling of the Fe K$\alpha$ emission line can be achieved using an
% X-ray reflection model such as \textsc{relxill} \citep{Garcia2014}, and we attempted spectral fits using \textsc{relxill}, \textsc{xillver} and \textsc{xillvercp} (flavours of the \textsc{relxill} model), finding that these models were not well suited to modelling the softer X-ray components of the spectrum. \textcolor{ForestGreen}{Nonetheless, \textsc{laor} is capable of modelling emission lines from the accretion discs around rotating compact objects, accounting for the origin of the line (inner or outer disc), the relativistic effects and if the line is single or double peaked. Therefore, it provides a reliable constraint on the line energy and disc inclination.}
There are 3 further components in our spectral model: \texttt{E}, \texttt{A} and \texttt{G}. Here, \texttt{E} corresponds to two absorption edges (\textsc{edge}) at $0.48$ keV and $2.22$ keV. These features likely arise from \textit{NICER} calibration systematics as they could be attributed to Oxygen ($\sim 0.5$ keV) and a gold M edge (2.1$-$4.5 keV complex) \citep{Wang2021}. \texttt{A} corresponds to three Gaussian absorption lines (\textsc{gabs}) at energies $2.37$ keV, $2.79$ keV and $6.97$ keV. The first two are likely associated with the aforementioned gold M absorption while the third physically corresponds to Fe XXVI K$\alpha$. Lastly, \texttt{G} represents four Gaussian emission lines (\textsc{gauss}) at energies $1.77$ keV (Si VIII K$\alpha$), $0.718$ keV (Fe XVIII L$\beta$), $1.47$ keV (Al K$\alpha$) and $2.10$ keV (P XIV K$\beta$). These are assumed to be real features with the exception of Al K$\alpha$ which may arise from \textit{NICER} calibration systematics \citep{Wang2021}.

\begin{table}
\begin{center}
\begin{tabular}{ c  c  c } 
\hline
Model Component & Parameter & Value \\
\hline
\textsc{tbabs} & $\rm{N}_H$ [10$^{22}$ cm$^{-2}$] & 0.233 $\pm^{0.002}_{0.002}$ \\ 
\hline
\textsc{diskbb} & T$_{\text{in}}$ [keV] & 1.186 $\pm^{0.006}_{0.002}$ \\ 
\hline
\textsc{bbody} & $kT$ [keV] & 0.356 $\pm^{0.003}_{0.003}$ \\ 
\hline
\vspace*{0.1cm}
\textsc{laor} & $E$ [keV] & 6.573 $\pm^{0.057}_{0.060}$ \\
\vspace*{0.1cm}
& $\Gamma$ & 2.376 $\pm^{0.177}_{0.177}$ \\ 
\vspace*{0.1cm}
& R$_{\rm{in}}$ [GM/c$^{2}$] & 4.892 $\pm^{0.636}_{0.853}$ \\
\vspace*{0.1cm}
& R$_{\rm{out}}$ [GM/c$^{2}$] & 400.0$~^{a}$\\
\vspace*{0.1cm}
& $i$ [deg] & 86.20 $\pm^{0.323}_{0.028}$ \\
\hline
\vspace*{0.1cm}
Absorption Edges (\texttt{E}) & $E_1$ [keV] & 2.224 $\pm^{0.001}_{0.001}$ \\ 
\vspace*{0.1cm}
\textsc{edge}& $\tau_{\rm{max, 1}}$ [keV] & 0.368 $\pm^{0.005}_{0.006}$ \\ 
\vspace*{0.1cm}
& $E_2$ [keV] & 0.482 $\pm^{0.043}_{0.059}$ \\ 
\vspace*{0.1cm}
& $\tau_{\rm{max, 2}}$ [keV] & 0.803 $\pm^{0.006}_{0.002}$ \\ 
\hline
\vspace*{0.1cm}
Absorption Lines (\texttt{A}) & $E_1$ [keV] & 2.369 $\pm^{0.003}_{0.004}$ \\ 
\vspace*{0.1cm}
\textsc{gabs} & $\sigma_{1}$ [keV] & 0.038 $\pm^{0.007}_{0.007}$ \\
\vspace*{0.1cm}
& $E_2$ [keV] & 2.797 $\pm^{0.006}_{0.007}$ \\ 
\vspace*{0.1cm}
& $\sigma_{2}$ [keV] & 0.048 $\pm^{0.010}_{0.011}$ \\
\vspace*{0.1cm}
& $E_3$ [keV] & 6.974 $\pm^{0.008}_{0.008}$ \\ 
\vspace*{0.1cm}
& $\sigma_{3}$ [keV] & 0.002 $\pm^{0.043}_{0.001}$ \\
\hline
\vspace*{0.1cm}
Emission Lines (\texttt{G}) & $E_1$ [keV] & 1.772 $\pm^{0.004}_{0.004}$ \\ 
\vspace*{0.1cm}
\textsc{gauss} & $\sigma_{1}$ [keV] & 0.042 $\pm^{0.007}_{0.004}$ \\
\vspace*{0.1cm}
& $E_2$ [keV] & 0.7181 $\pm^{0.004}_{0.004}$ \\ 
\vspace*{0.1cm}
& $\sigma_{2}$ [keV] & 0.006 $\pm^{0.012}_{0.001}$ \\
\vspace*{0.1cm}
& $E_3$ [keV] & 1.466 $\pm^{0.007}_{0.006}$ \\ 
\vspace*{0.1cm}
& $\sigma_{3}$ [keV] & 0.043 $\pm^{0.009}_{0.009}$ \\
\vspace*{0.1cm}
& $E_4$ [keV] & 2.071 $\pm^{0.007}_{0.007}$ \\ 
\vspace*{0.1cm}
& $\sigma_{4}$ [keV] & 0.016 $\pm^{0.019}_{0.002}$ \\
\hline
\end{tabular}
\caption{\label{tb:SpecFit} Best-fitting parameters from our fit to the time-averaged spectrum of \textit{Swift} J1858.6 -0814, achieving a reduced $\chi^2$ is $\chi^2/\nu=834.42/815$. The absorption edges at centroid energies $E_1$ and $E_2$ are likely features arising from \textit{NICER} calibration systematics, as they could be associated to oxygen and a gold M edge (2.1$-$4.5 keV complex) respectively \citep{Wang2021}. Gaussian absorption lines (\textsc{gabs}) at energies $2.37$ keV and $2.79$ keV also fall within this absorption complex. The (\textsc{gabs}) component at $6.97$ physically corresponds to Fe XXVI K$\alpha$. The four Gaussian emission lines are at energies $0.718$ keV (Fe XVIII L$\beta$), $1.47$ keV (Al K$\alpha$) , $1.77$ keV (Si VIII K$\alpha$), and $2.10$ keV (P XIV K$\beta$). These are assumed to be real features with the exception of Al K$\alpha$ which may also arise from \textit{NICER} calibration systematics.\\
$^{a}$ Parameter fixed for the duration of the fit.}
\end{center}
\end{table}

We show the resulting model (red) in Fig. \ref{fig:SpecFit}. The individual components, respectively, are coloured green, magenta and blue which correspond to the Laor iron line profile, the NS surface blackbody and the multi-temperature blackbody originating from the accretion disc. The astrophysical emission lines at $0.718$ keV (Fe L$\beta$), $1.77$ keV (Si K$\alpha$), and $2.10$ keV (P k$\beta$) are shown in cyan and the $1.47$ keV (Al K$\alpha$) emission line, suspected to be present as a result of \textit{NICER} calibration systematics is shown in orange. Our fit returns the parameters listed in Table \ref{tb:SpecFit}, a reduced $\chi^2$ of $\chi^2/\nu = 834.42/815$ and the corresponding null hypothesis probability is $p = 0.414$. The eclipse duration (the sum of the duration of the ingress, totality and egress) is $\sim 4400$ s, corresponding to $\sim 0.06 \%$ of the orbital period. Also, the in-eclipse count rate is low. As such, it is reasonable to assume the out-of-eclipse can be approximated by the time-averaged spectrum. Therefore, we use the time-averaged spectrum within the eclipse profile model. Note that the eclipse profile model is not sensitive to the X-ray spectral model used, but simply requires a reasonable fit to the observed spectrum (see \citealt{Knight2021}). Since we assume an X-ray point source, any spectral decomposition that fits the observed spectrum would have the same time dependence. Therefore, the X-ray spectral model is not critical to the results obtained via eclipse mapping. As such we are content with our spectral model including a \textsc{laor} iron line profile and we do not pursue a more complex reflection model.
%We achieve an acceptable fit with a reduced $\chi^2$ value of $\chi^2/\nu = 834.42/815$ and a null hypothesis probability, $p = 0.414$. The best-fitting parameters are reported in Table \ref{tb:SpecFit}. Since the eclipse duration (duration of ingress, totality and egress) is $< 0.06 \%$ of the orbital period and the in-eclipse count rate is low, we assume the time-averaged spectrum is approximately the out-of-eclipse spectrum and subsequently use this model within our eclipse profile model.

\section{Eclipse Mapping}
\label{sx:ecmodel}
In Section \ref{sx:data} we demonstrated that the observed eclipse profiles of \textit{Swift} J1858.6$-$0814 have many of the same characteristics (extended in time, asymmetric and energy-dependent) as those of EXO 0748$-$676 \citep{Parmar1991, Wolff2009, Knight2021}. Furthermore, an orbital phase-resolved spectral analysis of the ingress and egress regions in \textit{Swift} J1858.6-0814, returns similar results to our previous analysis of EXO 0748$-$676 \citep{Knight2021}. As such, it seems sensible to use the same modelling approach for \textit{Swift} J1858.6-0814 as we used for EXO 0748$-$676. We find that the abundance of similarities in the eclipse profiles of these two sources requires a thorough discussion, thus we separate our findings. We present the constraints on binary inclination, $i$ and mass ratio, $q$ obtained via our eclipse profile modelling here, and will present the results of our phase-resolved spectral analysis in a forthcoming study (Knight et al. in prep). Full details of the eclipse mapping model can be found in \cite{Knight2021}, however, we include a self-contained summary here. 

\subsection{Eclipse Profile Model}
\label{sx:Model}
We assume the companion star is spherically symmetric. The star itself is optically thick, but has a layer of optically thin absorbing material surrounding it. The eclipse transitions, therefore, start when a line-of-sight first passes through this absorbing layer and the hydrogen column density for the line-of-sight is \citep{Knight2021}:
%We model the companion star as spherically symmetric with radial hydrogen number density profile $n_\star (r)$. The photospheric radius of the star is $R_{\rm cs}$, such that $n_\star (r)$ in the region $r<R_{\rm cs}$ is large enough for the optical depth to be effectively infinite. We represent distances in units of the companion star radius, $x \equiv r / R_{\rm cs}$, and define the \textit{impact parameter}, $b(t)$, as the projected separation between NS and companion star in units of $R_{\rm cs}$. Therefore, totality occurs when $b(t) \leq 1$. For $b(t) > 1$, the hydrogen column density for a sight-line through the surrounding material is \citep{Knight2021}: 
\begin{equation}
    \rm{N}_H(t) = 2 N_{H,0} \int_{b(t)}^{x_{\rm out}} n(x) \frac{x}{\sqrt{x^2-b(t)^2}} dx,
    \label{eqn:Nht}
\end{equation}
%\cmtmm{should the integral really have a lower limit of b(t)?}\cmtai{yes}
where $n(x)$ 
%\equiv n_\star(r)/n_0$ such that $n_0=n_\star(r=R_{\rm cs})$
is the radial density profile of the surrounding material, $N_{\rm{H},0}$ is \textit{surface column density}, $x$ is the distance from the surface of the companion star defined in units of the companion star radius, $b(t)$ is the \textit{impact parameter} \citep{Knight2021} and
%=R_{\rm cs}n_0$ is the column density of a sight-line of length $R_{\rm cs}$ through material with constant density $n_0$. 
$x_{\rm out}$ represents the furthest distance from the companion star surface where the density of the material layer is non-negligible.
Assuming circular orbits, and that the companion is filling its Roche-Lobe, the inclination, $i$, and mass ratio, $q$ are related via the totality duration, $t_e$, and orbital period, $P$:
\begin{equation}
    \sin i = \frac{\sqrt{1-h^2(q)}}{ \cos(\pi t_e / P) }.
    \label{eqn:inc}
\end{equation}
where $h(q)$ is the ratio of the Roche-Lobe radius to the orbital separation (e.g. equation 8 from \citealt{Knight2021}). The impact parameter is a function of orbital phase, inclination and $h(q)$ (equation 6 from \citealt{Knight2021}).
Therefore, to calculate the impact parameter as a function of orbital phase, the only model parameter required is $q$ when $t_{\rm e}$ and $P$ are known.
%Therefore, for a known $t_{\rm e}$ and $P$, the only model parameter required to calculate $b$ as a function of orbital phase is $q$.
% Therefore, for known $t_{\rm e}$, $P$ and $t_0$, the only model parameter required to calculate $b(t)$ is $q$.
%The column density $\rm{N}_H(t)$ can be calculated from an assumed radial density function, $n(x)$, and normalisation $N_{\rm{H},0}$, with $q$ as the only other model parameter. 

Four radial density functions are currently provided within the eclipse profile model. These are
\begin{enumerate}
    \item a power-law with index $m$ corresponding to a stellar wind with constant velocity; $n(x) = x^{-m}$,
    \\
    \item an accelerating stellar wind with acceleration parameter $\beta$; $n(x) = x^{-2}~( 1 - x^{-1} )^{-\beta}$,
    \\
    \item a Gaussian outflow described by the fractional width of material, $\Delta$; $n(x) = \exp\left[ -\frac{(x-1)^2}{2\Delta^2} \right]$,
    \\
    \item and an exponential outflow described by material scale height, $h$; $n(x) = \exp \left[ \frac{1-x}{h} \right]$.
\end{enumerate}
Here i and ii are often seen in stellar wind modelling (e.g. \citealt{vanderHelm, Puls2008}) while iii and iv would be more typical of ablation or atmosphere modelling \citep{Knight2021}. Our model additionally allows the material layer to be turned off by instead applying an abrupt transition between out-of-eclipse and totality. Thus, the ingress and egress are modelled as straight lines.

The time-dependent specific photon flux, $S(E,t)$, is the product of an energy dependent transmission factor, $\alpha(E)$ and the out-of-eclipse spectrum, $S_0(E)$ (approximated by the time-averaged spectrum in Section \ref{sx:SpectralFit}). We use our absorption and scattering model \textsc{abssca} \citep{Knight2021} to calculate $\alpha(E)$, thus introducing the ionisation parameter ($\xi$) and covering fraction ($\rm{f}_{\rm cov}$) as properties of the absorbing material. These material properties are free model parameters and their ingress values ($\xi_{\rm in}$ and $\rm{f}_{\rm cov,in}$) can differ from their egress values ($\xi_{\rm eg}$ and $\rm{f}_{\rm cov,eg}$). All forms of $n(x)$ are trialled, in addition to trying no material layer. Each of the characteristic density function parameters ($m$, $\beta$, $\Delta$ or $h$, depending on the $n(x)$ form being used) can have different values during the ingress and the egress.
%We fix the orbital period to $P=21.3$ hr \citep{Buisson2021} and leave the totality duration $t_e$ as a free model parameter.

\subsection{Results}
\begin{table}
\begin{center}
\begin{tabular}{|c|c|c|c|c|c|} 
\hline
Density Profile & Parameter(s) & $\chi^{2}$ & $\nu$ & $p$ & $q$ \\
\hline
 No Material & - & 104600 & 2675 & - & - \\ 
 \hline
Power-law & m\ =\ 2.00$~^{a}$ & 99100 & 2675 & - & -\\
 & m\ =\ 10.0$~^{a}$ & 32900 & 2675 & - & - \\ 
 \\
 & $\rm{m_{in}}$ \ =\ 198.2 & 2696.0 & 2673 & $10^{-102}$ & 0.139\\ 
  & $\rm{m_{eg}}$ \ =\ 178.9 &  & & \\
\hline
Accelerating & $\beta_{\rm{in}}$\ =\ 6.12 & 4640.6 & 2673 & $10^{-136}$ & 0.137\\ 
 & $\beta_{\rm{eg}}$\ =\ 7.68 & & \\ 
\hline
Gaussian & $\Delta_{\rm{in}}$\ =\ 0.0099 & 2723.6 & 2673 & 0.433 & 0.140 \\ 
 & $\Delta_{\rm{eg}}$\ =\ 0.0119 & & \\ 
\hline
Exponential & $\rm{h}_{\rm{in}}$\ =\ 0.0086 & 2679.0 & 2673 & 0.464 & 0.139 \\ 
 & $\rm{h}_{\rm{eg}}$\ =\ 0.0135 & & \\ 
\hline \\
\end{tabular}
\end{center}
\vspace{-0.5cm}
\caption{\label{tb:fitstats} For each radial density profile, the characteristic density profile parameters and associated fit statistics are found by fitting the eclipse profile model to the eclipse profiles of \textit{Swift} J1858.6$-$0814 in multiple energy bands. The best-fitting values of the key density profile parameters (power-law index, $m$, acceleration parameter, $\beta$, fractional width of the material, $\Delta$, and fractional scale height, $h$, for the power-law, accelerating, Gaussian and exponential density profiles respectively) are each presented with their associated chi-squared, $\chi^{2}$, the number of degrees-of-freedom, $\nu$, the null hypothesis probability, $p$, and the mass ratio, $q$ for all fits where constraints were possible.\\
%Characteristic density profile parameters and corresponding fit statistics from fitting the eclipse profile model simultaneously to eclipse profiles of \textit{Swift} J1858.6$-$0814 in five energy bands, for each possible density profile. From left to right: the assumed density profile of the absorbing material, key parameters governing the density profile (power-law index, $m$, acceleration parameter, $\beta$, fractional width of the material, $\Delta$, and fractional scale height, $h$, for the power-law, accelerating, Gaussian and exponential density profiles respectively), chi-squared, number of degrees-of-freedom, null hypothesis probability, $p$, and mass ratio.
$^{a}$ Parameter fixed for the duration of the fit.
}
\end{table}

\begin{table}
\begin{center}
\begin{tabular}{|c|c|c|}
\hline
Parameter & Gaussian Density Profile & Exponential Density Profile\\
\hline
P [s] & 76841.3 & 76841.3\\
\hline
$t_{\rm e}$ & 4103.52 $\pm ^{1.19}_{1.03}$ & $4096.17 \pm ^{0.02}_{1.66}$\\
\hline
$q$ & 0.1402 $\pm ^{0.0028}_{0.0037}$ & 0.1394 $\pm ^{0.0021}_{0.0019}$ \\
\hline
$i^{\circ}$ & 80.90 $\pm ^{0.13}_{0.10}$ & 80.96 $\pm^{0.08}_{0.09}$ \\
\hline
$\log(\xi)_{\rm{in}}$ & 1.935 $\pm ^{0.010}_{0.007}$ & 1.906 $\pm ^{0.044}_{0.032}$\\
\hline
$\log(\xi)_{\rm{eg}}$ & 1.953 $\pm ^{0.016}_{0.004}$ & 1.910 $\pm ^{0.012}_{0.012}$ \\
\hline
$\rm{f_{cov, in}}$ & 0.999 $\pm _{0.0001}^{0.001}$ & 0.980 $\pm^{0.009}_{0.016}$ \\
\hline
$\rm{f_{cov, eg}}$ & 0.997 $\pm _{0.005}^{0.003}$ & 0.984 $\pm ^{0.009}_{0.005}$\\
\hline
$\Delta_{\rm{in}}$ or $\rm{h}_{\rm{in}}$ & 0.0099 $\pm _{0.0010}^{0.0004}$ & 0.0086 $\pm ^{0.00004}_{0.00110}$\\
\hline
$\Delta_{\rm{eg}}$ or $\rm{h}_{\rm{eg}}$  & 0.0119 $\pm _{0.0004}^{0.0003}$ & 0.0135 $\pm ^{0.00092}_{0.00004}$ \\
\hline
\\
\end{tabular}
\end{center}
\vspace{-0.5cm}
\caption{\label{tb:fitpars} Best-fitting model parameters, for both the Gaussian and exponential density profiles, found by fitting the eclipse profile model to the eclipse profiles of \textit{Swift} J1858.6$-$0814 in five energy bands simultaneously. The parameters are orbital period (fixed for the duration of the fitting), totality duration ($t_{e}$), mass ratio ($q$), binary inclination (subsequently calculated using the best-fitting $q$ and $t_{e}$), log of the ionisation parameter for the ingress and egress ($\log(\xi)_{\rm{in}}$ and $\log(\xi)_{\rm{eg}}$), covering fraction for the ingress and egress ($\rm{f_{cov, in}}$ and $\rm{f_{cov, eg}}$, and the characteristic density profile parameter for the ingress and egress. These characteristic parameters are the fractional widths of the material layer ($\Delta$) for the Gaussian model and material scale heights (h) for the exponential model. Values are provided with a 1 $\sigma$ error obtained via MCMC. 
}
\end{table}

\begin{figure*}
\centering
\includegraphics[width=0.8\textwidth]{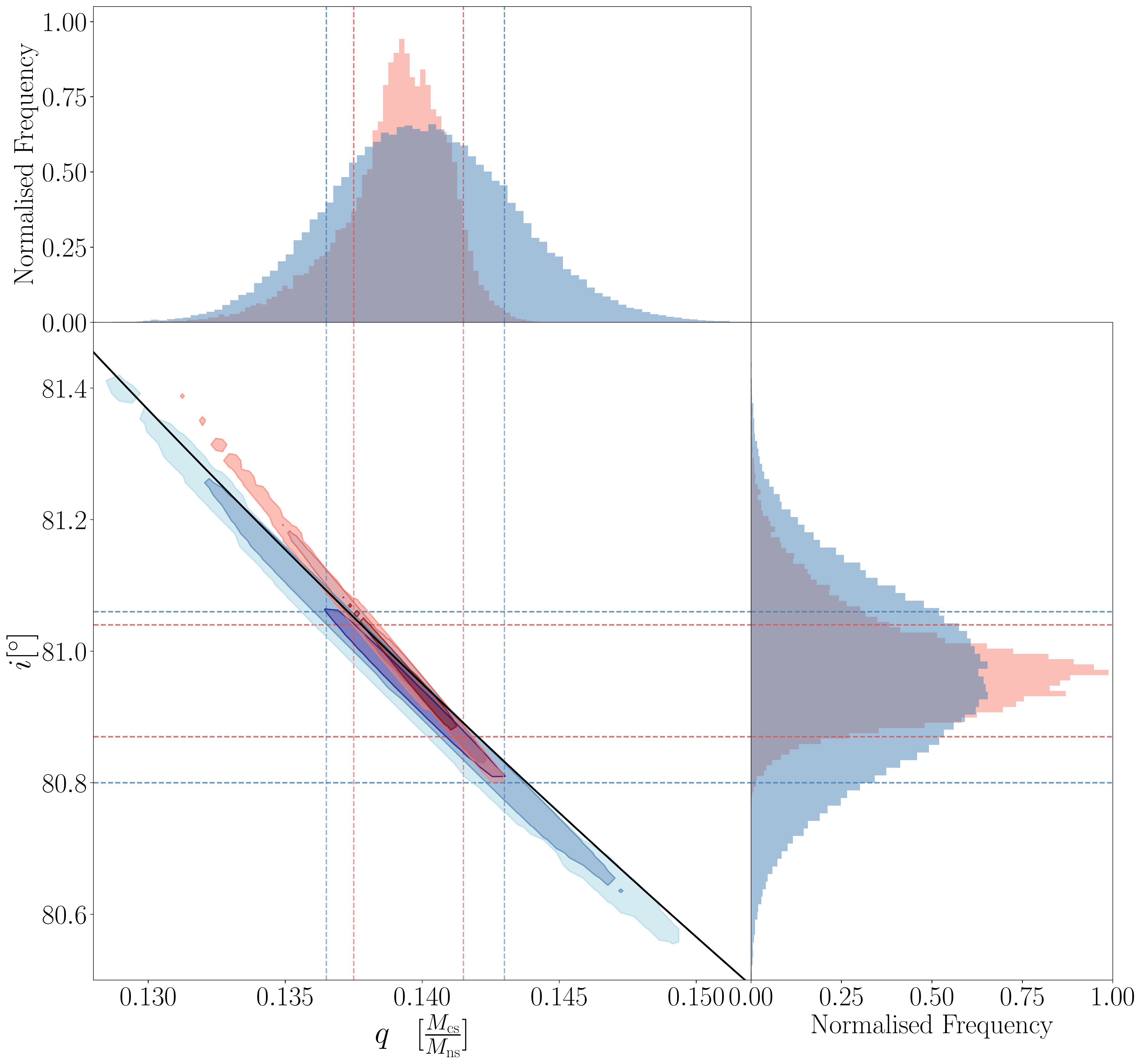}
\vspace{-0.25cm}
\caption{Posterior distributions for the mass ratio, $q$ (top), and binary inclination, $i$ (right). These are shown for both the Gaussian model (blue) and exponential model (red). The distributions are obtained by running a Markov Chain Monte Carlo simulation with 256 walkers, 768000 steps and a burn-in length of 742912. Blue and red dashed lines show $1 \sigma$ confidence intervals for the Gaussian and exponential models respectively. The centre plot shows a 2D projection of these posterior distributions for both density profile models, plotted with the theoretical $q - i$ relation (black). Dark, medium and light shades of blue and red highlight $1$, $2$ and $3 \sigma$ contours in this 2D parameter space.}
\label{fig:dist}
\end{figure*}

We simultaneously fit the eclipse profiles of \textit{Swift} J1858.6 -0814 in five energy bands; $0.4 - 1.0$ keV, $1.0 - 2.0$ keV, $2.0 - 4.0$ keV, $4.0 - 6.0$ keV and $6.0 - 10.0$ keV using \textsc{xspec} v12.12.0. All eclipse profiles are obtained following the procedure described in Section \ref{sx:EcProf}. Note the $6.0 - 10.0$ keV band is preferred to separate $6.0 - 8.0$ keV and $8.0 - 10.0$ keV bands as the former has a higher number of counts per bin, thus allowing the use of chi-squared fit statistics. We ensure the best-fitting model parameters primarily correspond to the eclipse transitions, which host the energy-dependent behaviour, by ignoring most of the out-of-eclipse and totality phase bins. Despite this, a systematic error of $15$ per cent is applied to account for the variability in the remaining out-of-eclipse portion of the data. The eclipse duration has previously been measured as $t_e = 4098 \pm ^{17}_{18}$ s \citep{Buisson2021}, therefore, we initially set $t_e = 4100$ s but keep it as an unconstrained free parameter during the fits. 

Following the modelling approach of \citet{Knight2021}, we first trial our eclipse profile model with no absorbing material surrounding the companion. Given the extended ingress and egress duration in the eclipse profiles, it is unsurprising that this model fits the data poorly ($\chi^{2} / \nu = 104600 / 2675$). Therefore, the data require some absorbing medium surrounding the companion star to recreate the extended and asymmetric ingress and egress. We test the four radial density profiles detailed in Section \ref{sx:Model} and present the resulting fit statistics in Table \ref{tb:fitstats}. Although the assumed form of the density profile has a large influence on fit quality, it has little effect on the inferred mass ratio, which is $q\sim 0.14$ for all fits where constraints are possible.

Assuming reasonable power-law indices of $m = 2$ and $m = 10$, the power-law radial density profile does not yield a good fit to the data. Respectively, the reduced chi-squared are  $\chi^{2} / \nu = 99100 / 2675$ and $\chi^{2} / \nu = 32900 / 2675$, the associated null-hypothesis probabilities are negligible and the mass ratios are unconstrained.
%The power-law density profile with reasonable power-law indices of $m = 2$ and $m = 10$, are unable to describe the data, yielding $\chi^{2} / \nu = 99100 / 2675$ and $\chi^{2} / \nu = 32900 / 2675$ respectively. The corresponding null-hypothesis probability is negligible and the mass ratio unconstrained. 
Allowing the power-law indices to be free during the fits yields unphysical values of $m_{\rm{in}} = 198.2$ and $m_{\rm{eg}} = 178.9$ for the ingress and egress respectively. While a statically better fit is achieved here ($\chi^{2} / \nu = 2969.0 / 2673 $) than for the fits assuming power-law indices of $m = 2.0$ and $m = 10.0$, it is clear that the data require a steeper radial density function. Thus, we discard the power-law density profile. 

The accelerating wind profile is a steeper function, with the density dropping off more quickly with distance from the companion star's surface, thus could improve upon the fits using the power-law density profile. However, this radial density function is simply too steep to model the heavily extended eclipse transitions we observe and is, therefore, discarded. The best-fitting acceleration parameters for the ingress and egress respectively are $\beta_{\rm{in}} = 6.12$ and $\beta_{\rm{eg}} = 7.68$, yielding $\chi^{2} / \nu = 4640.6 / 2673$, and the associated null-hypothesis probability is, $p = 10^{-136}$.

The remaining two density profiles yield acceptable fits to the observed eclipse profiles. We show the resulting eclipse profiles for both the Gaussian and exponential density profiles in Fig. \ref{fig:ecfits}A(i) - A(iv) and Fig. \ref{fig:ecfits}B(i) - B(iv) respectively. For the Gaussian and exponential density profiles we obtain $\chi^{2} / \nu = 2723.6 / 2673 $ ($p = 0.433$) and $\chi^{2} / \nu = 2679.0 / 2673 $  ($p = 0.464$) respectively. Since both fits are statistically similar, we consider both density profiles in our subsequent analysis and present the best fitting parameters from both models in Table \ref{tb:fitpars}. 

The best-fitting model parameters obtained assuming the Gaussian radial density profile are found to be consistent with the best-fitting model parameters obtained assuming the exponential radial density profile within a $1 \sigma$ error, thus increasing confidence in our constraints on $q$ and subsequent constraints on $i$. We find consistent ionisation parameters between the ingress and egress side of the star, which is in contrast to EXO 0748$-$676 where the ingress appeared to be more heavily ionised \citep{Knight2021}. Overall, the surrounding material is less ionised in \textit{Swift} J1858.6$-$0814 where log$(\xi) \sim 1.9$ than for EXO 0748$-$676 where log$(\xi) \sim 3.0$. We further find consistent covering fractions between the ingress and egress, in contrast to EXO 0748$-$676, where the Gaussian model suggested the leading side of the companion to be less covered. The energy-dependent eclipse timings predicted by the model strongly depend on the density profile chosen, since this governs the X-ray absorption in the surrounding medium. We find both density profiles can recreate both t$_{90}$ and t$_{10}$ behaviour presented in Fig. \ref{fig:T90}. The t$_{10}$ behaviour is more difficult to model because it relies on a sufficiently high material density and a high material ionisation. Nonetheless, our modelling can reproduce this and the overall eclipse models are shown in Fig. \ref{fig:ecfits}. 

\subsection{Binary Inclination and Mass Ratio}
Our fits to the observed eclipse profiles assuming a Gaussian and exponential density profile each return best-fitting values for model parameters $t_e$ and $q$. The binary inclination $i$ can then be found from these parameters and Equation \ref{eqn:inc}. We constrain posterior distributions of $i$ and $q$ by running an Markov Chain Monte Carlo (MCMC) simulation within \textsc{xspec} using 256 walkers, a total of 768000 steps and a burn-in period of 742912 steps (see Appendix \ref{sx:ap1} for further details). For each step in the chain, we calculate $i$ from $q$ and $t_e$ and present the resulting posterior distribution in Fig. \ref{fig:dist} (side panel), in which the Gaussian and exponential models are coloured blue and red respectively. Corresponding $1 \sigma$ contours are provided by the dashed lines of the same colours. 

Our posterior distributions demonstrate tight constraints on both $q$ and $i$ finding, at $1 \sigma$, $q = 0.1402 \pm ^{0.0028}_{0.0037}$ and $i = 80.9^{\circ} \pm ^{0.13}_{0.10}$ when assuming Gaussian density profile and $q = 0.1394 \pm ^{0.0021}_{0.0019}$ and $i = 80.96^{\circ} \pm ^{0.08}_{0.09}$ when assuming the exponential density profile. We further demonstrate this in Fig. \ref{fig:dist} where 1, 2 and $3 \sigma$ regions of $q - i$ parameter space are shaded in light, medium and dark shades of blue and red for the Gaussian and exponential models respectively. The black line shows the relation between $q$ and $i$ assuming a totality duration of $t_e = 4100$ s. We see that the fit closely follows this line, but width is introduced into the 2D contour by statistical uncertainty on $t_e$. We can, therefore, combine these results with future radial velocity amplitude measurements for a precision NS mass measurement.

We note the 5 degree discrepancy between the inclination measured via eclipse mapping ($\sim 81^{\circ}$) and the preferred inclination of the \textsc{laor} iron line profile used in our spectral fitting ($\sim 86^{\circ}$). This difference is very small when considering the modelling uncertainties and we consider their similarity encouraging. Although, eclipse mapping is measuring the inclination of the binary orbit itself while the \textsc{laor} model is measuring the inclination closer to the compact object. This could indicate that the inner disc is misaligned with the binary plane, but likely by a reasonably small amount (e.g. \citealt{Fragos2010}). 
%\cmtmm{are there strong QPOs in this source?}\cmtai{Good question, I'm not sure.} \cmtak{One QPO has been found at $\sim 10$ mHz - \citealt{Buisson2020}}

\subsection{Nature of the Companion Star}

\begin{figure}
\centering
\includegraphics[width=\columnwidth]{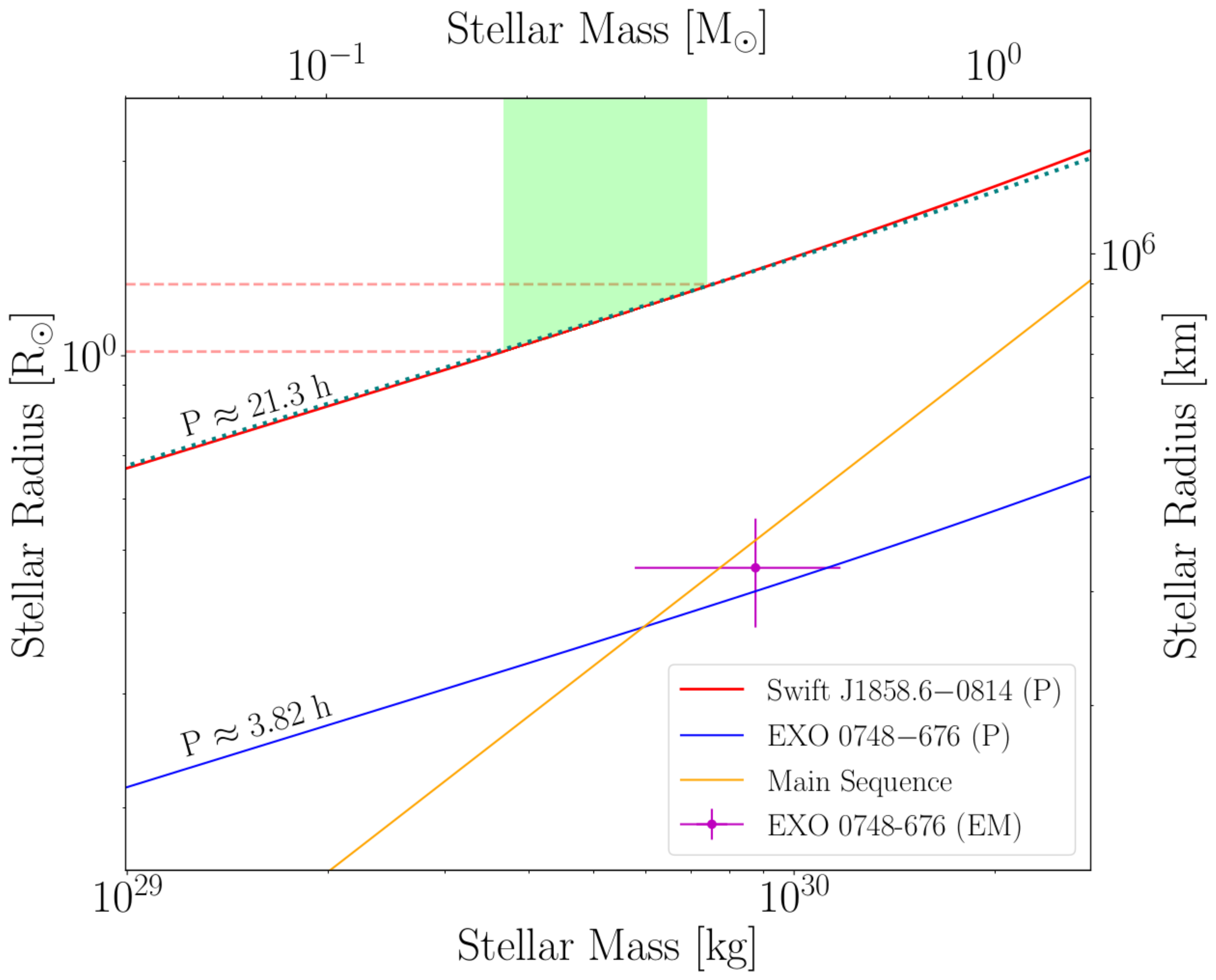}
\vspace{-0.5cm}
\caption{The mass-radius (M-R) relation for \textit{Swift} J1858.6$-$0814 derived using the orbital period of 21.3 hrs, a range of mass ratios, and assuming the companion star fills its Roche lobe (e.g \citealt{Buisson2021}). We show this for both $M_{\rm{ns}} = 1.4 M_{\odot}$ (solid, red) and $M_{\rm{ns}} = 2.5 M_{\odot}$ (dotted, teal). We follow the same procedure to calculate the M-R relation for EXO 0748$-$676 (blue), assuming the orbital period of 3.82 hrs. For comparison, the main sequence M-R relation (yellow) from \citealt{Demircan1991} is provided. The companion in EXO 0748$-$676 is consistent with being on the main sequence but the companion in \textit{Swift} J1858.6$-$0814 is clearly inconsistent with being on the main sequence. The green band shows the range of possible companion star masses for \textit{Swift} J1858.6$-$0814, found using the $3 \sigma$ range of mass ratios ($0.131 \leq q \leq 0.149$) and $1.4 \leq M_{\rm{ns}} \leq 2.5 M_{\odot}$. The dashed red lines highlight the range of corresponding stellar radii. The companion in \textit{Swift} J1858.6$-$0814 therefore has a radius much larger than a main sequence star of the same mass.
}
\label{fig:Mass}
\end{figure}

We can make inferences about the companion star properties from the results of our model fits. The mass and radius can be constrained from our measured value of $q$. Given the (current) absence of a binary mass function, the mass estimate, $M_{\rm cs} = q M_{\rm ns}$, comes simply from assuming some reasonable range of possible NS masses. From Kepler's law, the radius and mass are related via the orbital period as \citep{Buisson2021}
\begin{equation}
R_{\rm cs} = h(q) \left[ \frac{G(M_{\rm ns}+M_{\rm cs}) P^2}{(2\pi)^2} \right]^{1/3}.
\end{equation}
The solid red and dotted teal lines in Fig. \ref{fig:Mass} show this relation for the orbital period of \textit{Swift} J1858$-$0814, assuming respectively $M_{\rm ns} = 1.4~M_\odot$ and $M_{\rm ns} = 2.5~M_\odot$. We see that the assumed NS mass has little influence on the relation. The green shaded area shows the range of $M_{\rm cs}$ values corresponding to $1.4~M_\odot \leq M_{\rm ns} \leq 2.5~M_\odot$ and our measured $3\sigma$ contour on $q$ (we use the distribution from our Gaussian model, which has larger uncertainties than the exponential model). This range of $0.183~M_\odot \leq M_{\rm cs} \leq 0.372 M_\odot$ corresponds to $1.02 ~R_\odot \leq R_{\rm cs} \leq 1.29 ~R_\odot$ (dashed red lines). Following \citet[][see their Fig. 9]{Buisson2021}, we also plot the theoretical mass-radius relation for an isolated main sequence star \citep{Demircan1991} as a solid yellow line. It is clear that the companion star has a radius much larger than a main sequence star of the same mass. This was also noted by \citet{Buisson2021}, but is now definitively confirmed by our measurement of the mass ratio. For comparison, we also plot the same relation for the orbital period of EXO 0748$-$676 (solid blue line) as well as an eclipse mapping measurement of $R_{\rm cs}$ and $M_{\rm cs}$ \citep{Knight2021} for that source utilizing the known binary mass function (magenta cross). Interestingly the EXO 0748$-$676 companion \textit{is} consistent with being on the main sequence.

Evidence for irradiation driven ablation of the companion star’s outer layers is found through the requirement of an additional layer of absorbing material surrounding the companion star beyond the Roche lobe radius. This layer is modelled such that the material’s density decreases with distance from the companion's surface, thus explaining the observed, extended and energy-dependent eclipses. The best-fitting parameters from our modelling, regardless of the assumed radial density profile, suggests this layer is asymmetric. In the case of the exponential density profile, our modelling yields scale heights of $h_{\rm{in}} = 0.0086$ and $h_{\rm{eg}} = 0.0135$ for the ingress and egress respectively. For the Gaussian density profile we obtain $\Delta_{\rm{in}} = 0.0099$ $\Delta_{\rm{eg}} = 0.0119$ for the ingress and egress respectively. This asymmetry is required by the data since the egress duration is more than $1.5$ times the ingress duration. This can be understood if the material layer is elongated behind the companion star due to its orbital motion in a diffuse ambient medium. 

In order to properly compare the height of the material layer around the companion predicted by the Gaussian and exponential models, we define a characteristic radius within which $68.27$ per cent of the mass of the layer is contained. For the Gaussian model, this is simply $y_0=\Delta$, and for the exponential model it is $y_0 = -h \ln(1- 0.6827)$. For the Gaussian and exponential density profiles, the characteristic size for the material on the leading side of the star is $y_{0} \approx 0.00999$ and $y_{0} \approx 0.00987$ respectively. For the egress side of the star these respectively increase by $\sim 1.21$ times to $y_{0} = 0.0119$ and by $\sim 1.57$ times to $y_{0} = 0.0155$. The characteristic size of the material layer on the ingress side of the companion inferred from the two density profiles are in remarkable agreement, both suggesting the size of the layer is $\sim 1 \%~R_{\rm cs}$. For the egress side of the companion, the characteristic size of the material layer differs between the two density profiles and lies within the range $1.2-1.56 \%~R_{\rm cs}$. Using the constraints on $R_{\rm cs}$ from the previous paragraph, this corresponds to a physical size in the range $\sim 8400 - 14000$ km for the trailing side of the companion and $\sim 7000 - 8900$ km for the leading side. 

\section{Discussion}
\label{sx:diss}

We have applied our previously published eclipse profile model \citep{Knight2021} to archival X-ray eclipses of \textit{Swift} J1858.6$-$0814 in multiple energy bands, from which we have measured a mass ratio of $q \sim 0.14$ and a binary inclination of $i \sim 81^{\circ}$. Assuming the NS mass to be in the range $1.4~M_\odot \leq M_{\rm{ns}} \leq 2.5~M_\odot$ indicates that the companion star has a low mass in the range $0.183 M_{\odot} \leq M_{\rm{cs}} \leq 0.372 M_{\odot}$ and a radius in the range $1.02 R_{\odot} \leq R_{\rm{cs}} \leq 1.29 R_{\odot}$ (see Fig. \ref{fig:Mass}). These radii are much larger than a main sequence star of the inferred mass. \citet{Buisson2021} concluded from similar arguments that the companion star is a sub-giant. Naively though, this seems unlikely since the sub-giant phase is a short-lived stage of stellar-evolution and eclipsing LMXBs are very rare, implying that it should be vanishingly unlikely for us to observe such a system.

The apparent low likelihood of the sub-giant scenario could, however, be counteracted by a selection effect. Specifically, if mass transfer is triggered by the expansion of the companion star as it evolves off the main sequence, then the likelihood of a given LMXB containing an evolved star becomes greater than the likelihood of observing an isolated star in an evolved state. Indeed, radio pulsars observed to be in circular orbits with low mass functions have been suggested to be LMXBs with sub-giant companions \citep{Verbunt1993}. Expanded, sub-giant companions have also been suggested to drive mass transfer in LMXBs with an orbital period in excess of $0.5$ days, e.g. Sco X-1 \citep{Gottlieb1975}. Since the orbital period of \textit{Swift} J1858.6$-$0814 is just under a day, it fits into this class of systems.
%Evolution and consequent expansion of the companion star 

However, the expected main sequence lifetime of an isolated $\sim 0.3 M_{\odot}$ star far exceeds a Hubble time, so the companion star simply wouldn't have yet evolved onto the giant branch unless its evolutionary path had been altered by binary interactions. We also note that the orbital period is short. Under the reasonable assumption that the NS evolved from an intermediate-high mass progenitor, the original orbital period would have been much larger. Therefore, the binary's evolution requires a mechanism to decrease the orbital separation while keeping the binary system intact. The latter could be avoided if the system formed via capture during a close encounter, as this allows the two components to evolve separately before forming a binary. A close encounter within a globular cluster is one of the formation scenarios considered for Sco X-1 \citep{Mirabel2003}. Like \textit{Swift} J1858.6$-$0814, Sco X-1 has a low mass companion of $\sim 0.4 M_{\odot}$ \citep{Steeghs2002} and a similar orbital period of $\sim 18.9$ hrs \citep{Gottlieb1975}. Such a scenario would require \textit{Swift} J1858.6$-$0814 to have a high proper motion from being kicked out of the globular cluster in which it formed.

It seems more likely that the system instead had its separation reduced by a common envelope phase (CE; see \citealt{PPBinary} for a brief overview). This could have occurred prior to the formation of the NS if the intermediate-high mass progenitor overfilled its Roche lobe and led the system into a period of unstable mass transfer. Alternatively, the progenitor to the current companion could have been initially more massive, initiating the CE itself. Regardless of when the CE occurred, the system would have ejected mass and angular momentum during the CE phase, subsequently forming a short period binary.

An evolutionary path similar to that suggested for PSR J1952+2630 by \citet{Lazarus2013} seems plausible here if both binary components were initially of intermediate mass. This scenario assumes the binary components evolved together and that the binary remained intact after the formation of the NS. Subsequently, there is a period of mass transfer as an intermediate-mass X-ray binary (IMXB), during which the secondary loses some mass through accretion and ablation processes. Such an IMXB phase has also been suggested to have occurred during the evolution of Sco X-1 \citep{Chen2017}. Towards the end of the IMXB phase, Roche lobe overflow can lead to a period of dynamically unstable mass transfer and the creation of a CE \citep{Lazarus2013}. This assumes the intermediate-mass companion is on the tip of the red-giant or asymptotic-giant branch and the binary has a wide orbit. The wide initial orbit allows the system to survive the CE, emerging as a short period binary consisting of a NS and a stripped He star.

The idea that the companion is a stripped He star is intriguing. Its formation through CE  provides a means to significantly reduce the mass of the companion. Additionally, stripped helium stars are suggested to expand to giant dimensions as a result of a continuously growing shell \citep{Dewi2002, Dewi2003, Yoon2012, Laplace2020}. This picture could therefore reconcile both the inferred low mass and large radius of the companion in \textit{Swift} J1858.6$-$0814 if the expanding shell of the He star due to the onset of shell He burning was the trigger of the 2018-2020 outburst. This scenario can be tested by using spectroscopy to search for evidence of CNO enhancement of the companion.

Regardless of the prior evolution of \textit{Swift} J1858.6$-$0814, our modelling requires the presence of an ionized layer of material around the companion star and is likely driven by X-ray ablation of the stellar surface. The inferred properties of this layer are very similar to those we inferred for EXO 0748$-$676, for which irradiation driven ablation was also the expected origin \citep{Knight2021}. X-ray ablation impacts the outermost layers of the stellar surface. The incident radiation from the NS and disc (see \citealt{CastroSegura2022} for discussion on the disc wind in \textit{Swift} J1858.6$-$0814) bombard the companion, liberating material from its surface that builds up around the star. The radial profile of this collected material is what we measure in our modelling. The overall result of ablation is mass loss from the companion star in addition to that lost via Roche Lobe overflow. Ablation is not expected to be efficient enough to substantially reduce the companion's mass (e.g \citealt{Ginzburg2020}), so additional factors such as accretion or CE ejection are likely required to explain the extremely low inferred companion mass in \textit{Swift} J1858.6$-$0814.

Ablation can, however, enhance the mass loss from the companion in LMXBs \citep{PPBinary}. The incident irradiation causing ablation can also induce other effects on the companion star, changing the expected evolution of the system. As discussed by \cite{Podsiadlowski1991a}, these are irradiation driven winds and irradiation driven expansion. If the envelope of the star is sufficiently irradiated by the incident X-ray flux, the star will try to expand by a factor of 2 to 4 in order to reach a new state of thermal equilibrium. This occurs as irradiation changes the degree of ionisation in the outer layers of the star and thus changes the star's effective surface boundary condition. \cite{Podsiadlowski1991a} shows that stellar expansion arising from X-ray irradiation is a function of the initial stellar radius and the incident X-ray flux. Using their calculations, we can infer that a $0.2 - 0.4 M_{\odot}$ main sequence companion would have a radius $\approx 0.6 - 1.0 R_{\odot}$ for an irradiating flux of $\log(F / \rm{erg s}^{-1} \rm{cm}^{-2}) \approx 11.6$. The X-ray flux incident on the companion from the NS can be estimated through the ratio $F_{x} / F_{\rm{det}} \approx D^{2} / r^{2}$, where $F_{x}$ and $F_{\rm{det}}$ are the X-ray flux incident on the companion star from the NS and the X-ray flux incident on our detector respectively. The orbital separation is $r$ and the distance to the source is $D$. The average X-ray flux incident on the detector is $F_{\rm{det}} \approx 7.4 \times 10^{-11}$ ergs/s/cm$^{2}$, which is calculated from the observed X-ray flux from 10 epochs as reported by \cite{VanDenEijnden2020}. The orbital separation is found through Kepler's law and the and assuming the distance to the source is $D = 13$ kpc \citep{Buisson2020} we find $\log(F_{x} / \rm{erg s}^{-1} \rm{cm}^{-2}) \sim 12.1$. Thus it appears plausible that the irradiation is driving the ablation and evolution of the companion star and could, therefore, be the origin of the inferred material layer around the companion. However, we note that the X-rays driving ablation are powered by accretion, therefore expansion via ablation cannot be the \textit{cause} of Roche Lobe overflow, they can simply further increase the size of the companion once Roche Lobe overflow has already begun.

While there are many unknowns regarding the prior evolution of \textit{Swift} J1858.6$-$0814, and we cannot favour any particular evolutionary scenario, we note that the \textit{irradiation scenario} is capable of explaining the origin of the surrounding material layer, the under-massive companion and its larger radius. Despite this, we note that the inferred mass and radius of the companion are somewhat extreme, and additional evidence of X-ray irradiation is required to support this conclusion. In addition, we consider it likely that a prior CE phase occurred if the current binary components evolved together, thus explaining the origin of the short binary period and providing a route for substantial mass loss from the system. Future spectroscopic studies could uncover evidence of CNO enhancement of the companion, thus providing support for a prior CE phase and assist in distinguishing between the possible formation scenarios.

\begin{figure*}
\centering
\includegraphics[width=\textwidth]{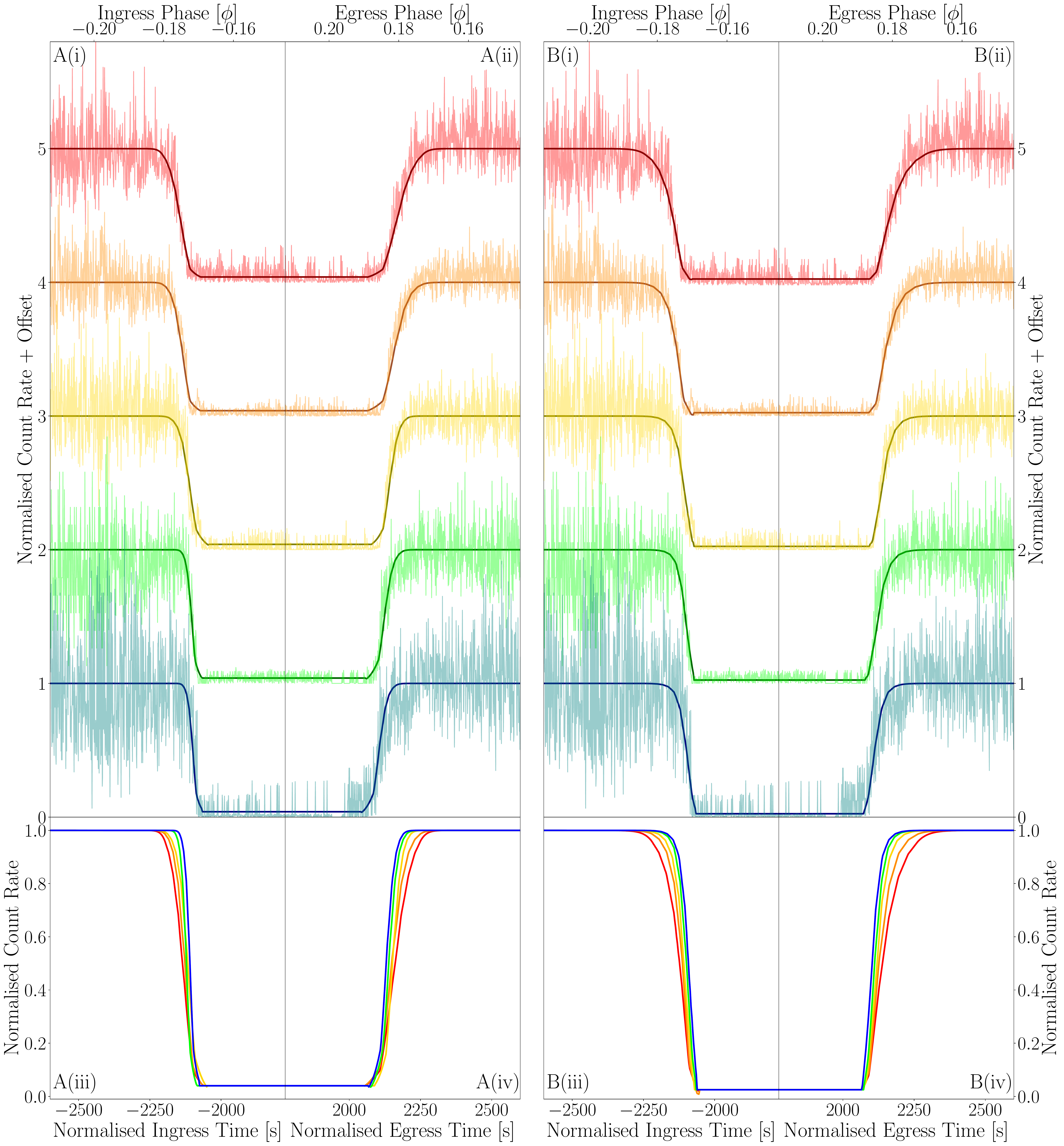}
\vspace{-0.5cm}
\caption{Resulting eclipse profiles obtained by simultaneously fitting the eclipse profiles of \textit{Swift} J1858.6$-$0814 in five energy bands ($0.4-1.0$ kev: red, $1.0-2.0$ kev: orange, $2.0-4.0$ kev: yellow, $4.0-6.0$ kev: green and $6.0-10.0$ kev: teal) with the eclipse profile model assuming the Gaussian radial density profile (A(i$-$iv)) and the exponential radial density profile (B(i$-$iv)). Panels A(i$-$ii) and B(i$-$ii): the resulting fits to each individual energy band, displayed with vertical offsets ($+0.0$ (teal), $+1.0$ (green), $+2.0$ (yellow), $+3.0$ (orange) and $+4.0$ (red)). Panels A(iii$-$iv) and B(iii$-$iv): the resulting eclipse profiles without a vertical offset, thus clearly displaying the energy dependent behaviour. For the Gaussian (A) and exponential (B) models respectively, $\chi^{2} / \nu = 2723.6 / 2673$ and $\chi^{2} / \nu = 2679.0 / 2673$.}
\label{fig:ecfits}
\end{figure*}

\section{Conclusions}
\label{sx:conclude}
We model the energy-dependent eclipse profiles of \textit{Swift} J1858.6$-$0814 in multiple energy bands, placing constraints on the binary inclination, $i$ and mass ratio, $q$. We find $i \sim 81 ^{\circ}$, and $q \sim 0.14$ which are related by the duration of totality, $t_e \sim 4100$ s. We combine our measured mass ratio with NS masses in the range $1.4~M_\odot \leq M_{\rm{ns}} \leq 2.5~M_\odot$ to infer that the companion star has a low mass in the range $0.183 M_{\odot} \leq M_{\rm{cs}} \leq 0.372 M_{\odot}$ and a large radius in the range $1.02 R_{\odot} \leq R_{\rm{cs}} \leq 1.29 R_{\odot}$. Since an isolated star with a mass in the inferred range would have main sequence lifetime in excess of the Hubble time, the large radius likely arises from of binary interactions. 

We consider it likely that a prior CE phase contributed to the ejection of mass from the system and the reduction in the orbital period, thus forming a short period binary with a low mass companion. If the companion emerged from the CE as a stripped star, it may swell to giant dimensions during later evolutionary stages. Future spectroscopic studies could confirm this possibility. An alternative scenario invokes irradiation of the companion star by the X-ray source, causing the companion star to expand (by a factor of 2-4 for low mass stars) to reach a new state of thermal equilibrium. The incident irradiation can also lead to enhanced mass loss \citep{Podsiadlowski1991a}. This scenario also explains the origin of the material layer found to surround the companion star by invoking irradiation driven ablation of the stellar surface. The inferred material layer is ionised and asymmetric ($21 - 57 \%$ thicker on the trailing side of the star than the leading side). This material layer is required by our eclipse profile model to recreate the observed extended and asymmetric eclipses in \textit{Swift} J1858.6$-$0814.

The extended and asymmetric eclipses in \textit{Swift} J1858.6$-$0814 are among numerous similarities between \textit{Swift} J1858.6$-$0814 and EXO 0748$-$676. We suggest that in both sources, the companion stars are being ablated by X-ray irradiation from the NS and disc. We will discuss these similarities in detail in a forthcoming study. 

\section*{Acknowledgements}
A. K. acknowledges support from the Oxford Hintze Centre for
Astrophysical Surveys, which is funded through generous support from the Hintze Family Charitable Foundation. A. I. acknowledges support from the Royal Society.

\section*{Data Availability}
The data used in this study are publicly available from the HEASARC website. The eclipse profile model is available upon reasonable request to the authors.

%%%%%%%%%%%%%%%%%%%%%%%%%%%%%%%%%%%%%%%%%%%%%%%%%%

%%%%%%%%%%%%%%%%%%%% REFERENCES %%%%%%%%%%%%%%%%%%

% The best way to enter references is to use BibTeX:

\bibliographystyle{mnras}
\bibliography{All_Refs} % if your bibtex file is called example.bib

\begin{thebibliography}{}
\makeatletter
\relax
\def\mn@urlcharsother{\let\do\@makeother \do\$\do\&\do\#\do\^\do\_\do\%\do\~}
\def\mn@doi{\begingroup\mn@urlcharsother \@ifnextchar [ {\mn@doi@}
  {\mn@doi@[]}}
\def\mn@doi@[#1]#2{\def\@tempa{#1}\ifx\@tempa\@empty \href
  {http://dx.doi.org/#2} {doi:#2}\else \href {http://dx.doi.org/#2} {#1}\fi
  \endgroup}
\def\mn@eprint#1#2{\mn@eprint@#1:#2::\@nil}
\def\mn@eprint@arXiv#1{\href {http://arxiv.org/abs/#1} {{\tt arXiv:#1}}}
\def\mn@eprint@dblp#1{\href {http://dblp.uni-trier.de/rec/bibtex/#1.xml}
  {dblp:#1}}
\def\mn@eprint@#1:#2:#3:#4\@nil{\def\@tempa {#1}\def\@tempb {#2}\def\@tempc
  {#3}\ifx \@tempc \@empty \let \@tempc \@tempb \let \@tempb \@tempa \fi \ifx
  \@tempb \@empty \def\@tempb {arXiv}\fi \@ifundefined
  {mn@eprint@\@tempb}{\@tempb:\@tempc}{\expandafter \expandafter \csname
  mn@eprint@\@tempb\endcsname \expandafter{\@tempc}}}

\bibitem[\protect\citeauthoryear{{Arnaud}}{{Arnaud}}{1996}]{Arnaud1996}
{Arnaud} K.~A.,  1996, in {Jacoby} G.~H.,  {Barnes} J.,  eds,  Astronomical
  Society of the Pacific Conference Series Vol. 101, Astronomical Data Analysis
  Software and Systems V. p.~17

\bibitem[\protect\citeauthoryear{{Baglio}, {Russell}, {Pirbhoy}, {Bahramian},
  {Heinke}, {Roche}  \& {Lewis}}{{Baglio} et~al.}{2018}]{Baglio2019}
{Baglio} M.~C.,  {Russell} D.~M.,  {Pirbhoy} S.,  {Bahramian} A.,  {Heinke}
  C.~O.,  {Roche} P.,   {Lewis} F.,  2018, The Astronomer's Telegram, \href
  {https://ui.adsabs.harvard.edu/abs/2018ATel12180....1B} {12180, 1}

\bibitem[\protect\citeauthoryear{Buisson et~al.,}{Buisson
  et~al.}{2020}]{Buisson2020}
Buisson D. J.~K.,  et~al., 2020, \mn@doi [Monthly Notices of the Royal
  Astronomical Society] {10.1093/mnras/staa2749}, 499, 793

\bibitem[\protect\citeauthoryear{Buisson et~al.,}{Buisson
  et~al.}{2021}]{Buisson2021}
Buisson D. J.~K.,  et~al., 2021, \mn@doi [Monthly Notices of the Royal
  Astronomical Society] {10.1093/mnras/stab863}, 503, 5600

\bibitem[\protect\citeauthoryear{Casares, Negueruela, Rib{\'{o}}, Ribas,
  Paredes, Herrero  \& Sim{\'{o}}n-D{\'{i}}az}{Casares
  et~al.}{2014}]{Casares2014}
Casares J.,  Negueruela I.,  Rib{\'{o}} M.,  Ribas I.,  Paredes J.~M.,  Herrero
  A.,   Sim{\'{o}}n-D{\'{i}}az S.,  2014, \mn@doi [Nature]
  {10.1038/nature12916}, 505, 378

\bibitem[\protect\citeauthoryear{Castro~Segura et~al.,}{Castro~Segura
  et~al.}{2022}]{CastroSegura2022}
Castro~Segura N.,  et~al., 2022, \mn@doi [Nature] {10.1038/s41586-021-04324-2},
  603, 52

\bibitem[\protect\citeauthoryear{Chen}{Chen}{2017}]{Chen2017}
Chen W.,  2017, \mn@doi [A\&A] {10.1051/0004-6361/201630239}, 606, A60

\bibitem[\protect\citeauthoryear{{Cominsky} \& {Wood}}{{Cominsky} \&
  {Wood}}{1984}]{Cominsky1984}
{Cominsky} L.~R.,  {Wood} K.~S.,  1984, \mn@doi [\apj] {10.1086/162361}, \href
  {https://ui.adsabs.harvard.edu/abs/1984ApJ...283..765C} {283, 765}

\bibitem[\protect\citeauthoryear{Demircan \& Kahraman}{Demircan \&
  Kahraman}{1991}]{Demircan1991}
Demircan O.,  Kahraman G.,  1991, \mn@doi [Astrophysics and Space Science]
  {10.1007/BF00639097}, 181, 313

\bibitem[\protect\citeauthoryear{{Dewi} \& {Pols}}{{Dewi} \&
  {Pols}}{2003}]{Dewi2003}
{Dewi} J.~D.~M.,  {Pols} O.~R.,  2003, \mn@doi [\mnras]
  {10.1046/j.1365-8711.2003.06844.x}, \href
  {https://ui.adsabs.harvard.edu/abs/2003MNRAS.344..629D} {344, 629}

\bibitem[\protect\citeauthoryear{{Dewi}, {Pols}, {Savonije}  \& {van den
  Heuvel}}{{Dewi} et~al.}{2002}]{Dewi2002}
{Dewi} J.~D.~M.,  {Pols} O.~R.,  {Savonije} G.~J.,   {van den Heuvel} E.~P.~J.,
   2002, \mn@doi [\mnras] {10.1046/j.1365-8711.2002.05257.x}, \href
  {https://ui.adsabs.harvard.edu/abs/2002MNRAS.331.1027D} {331, 1027}

\bibitem[\protect\citeauthoryear{Done}{Done}{2010}]{Done2010}
Done C.,  2010, \mn@doi [Accretion Processes In Astrophysics: IAC Winter School
  Of Astrophysics] {10.1017/CBO9781139343268.007}, 9781107030, 184

\bibitem[\protect\citeauthoryear{Dovciak}{Dovciak}{2004}]{Dovciak2004}
Dovciak M.,  2004, PhD Thesis

\bibitem[\protect\citeauthoryear{{Fabian}, {Rees}, {Stella}  \&
  {White}}{{Fabian} et~al.}{1989}]{Fabian1989}
{Fabian} A.~C.,  {Rees} M.~J.,  {Stella} L.,   {White} N.~E.,  1989, \mn@doi
  [\mnras] {10.1093/mnras/238.3.729}, \href
  {https://ui.adsabs.harvard.edu/abs/1989MNRAS.238..729F} {238, 729}

\bibitem[\protect\citeauthoryear{Fragos, Tremmel, Rantsiou  \&
  Belczynski}{Fragos et~al.}{2010}]{Fragos2010}
Fragos T.,  Tremmel M.,  Rantsiou E.,   Belczynski K.,  2010, \mn@doi [The
  Astrophysical Journal] {10.1088/2041-8205/719/1/l79}, 719, L79

\bibitem[\protect\citeauthoryear{{Garc{\'\i}a} et~al.,}{{Garc{\'\i}a}
  et~al.}{2014}]{Garcia2014}
{Garc{\'\i}a} J.,  et~al., 2014, \mn@doi [\apj] {10.1088/0004-637X/782/2/76},
  \href {https://ui.adsabs.harvard.edu/abs/2014ApJ...782...76G} {782, 76}

\bibitem[\protect\citeauthoryear{Ginzburg \& Quataert}{Ginzburg \&
  Quataert}{2020}]{Ginzburg2020}
Ginzburg S.,  Quataert E.,  2020, \mn@doi [Monthly Notices of the Royal
  Astronomical Society] {10.1093/mnras/staa3358}, 500, 1592

\bibitem[\protect\citeauthoryear{{Gottlieb}, {Wright}  \& {Liller}}{{Gottlieb}
  et~al.}{1975}]{Gottlieb1975}
{Gottlieb} E.~W.,  {Wright} E.~L.,   {Liller} W.,  1975, \mn@doi [\apjl]
  {10.1086/181703}, \href
  {https://ui.adsabs.harvard.edu/abs/1975ApJ...195L..33G} {195, L33}

\bibitem[\protect\citeauthoryear{Hare et~al.,}{Hare et~al.}{2020}]{Hare2020}
Hare J.,  et~al., 2020, \mn@doi [The Astrophysical Journal]
  {10.3847/1538-4357/ab6a12}, 890, 57

\bibitem[\protect\citeauthoryear{Higginbottom, Knigge, Long, Matthews, Sim  \&
  Hewitt}{Higginbottom et~al.}{2018}]{Higginbottom2018}
Higginbottom N.,  Knigge C.,  Long K.~S.,  Matthews J.~H.,  Sim S.~A.,   Hewitt
  H.~A.,  2018, \mn@doi [Monthly Notices of the Royal Astronomical Society]
  {10.1093/mnras/sty1599}, 479, 3651

\bibitem[\protect\citeauthoryear{{Hjellming} \& {Johnston}}{{Hjellming} \&
  {Johnston}}{1981}]{Hjellming1981}
{Hjellming} R.~M.,  {Johnston} K.~J.,  1981, \mn@doi [\apjl] {10.1086/183571},
  \href {https://ui.adsabs.harvard.edu/abs/1981ApJ...246L.141H} {246, L141}

\bibitem[\protect\citeauthoryear{Horne}{Horne}{1985}]{Horne1985}
Horne K.,  1985, \mn@doi [Monthly Notices of the Royal Astronomical Society]
  {10.1093/mnras/213.2.129}, 213, 129

\bibitem[\protect\citeauthoryear{Kaaret \& Ford}{Kaaret \&
  Ford}{1997}]{Kaaret1997}
Kaaret P.,  Ford E.~C.,  1997, \mn@doi [Science]
  {10.1126/science.276.5317.1386}, 276, 1386

\bibitem[\protect\citeauthoryear{Knight, Ingram, Middleton  \& Drake}{Knight
  et~al.}{2021}]{Knight2021}
Knight A.~H.,  Ingram A.,  Middleton M.,   Drake J.,  2021, \mn@doi [Monthly
  Notices of the Royal Astronomical Society] {10.1093/mnras/stab3722}, 510,
  4736

\bibitem[\protect\citeauthoryear{{Krimm} et~al.,}{{Krimm}
  et~al.}{2018}]{Krimm2018}
{Krimm} H.~A.,  et~al., 2018, The Astronomer's Telegram, \href
  {https://ui.adsabs.harvard.edu/abs/2018ATel12151....1K} {12151, 1}

\bibitem[\protect\citeauthoryear{{Laor}}{{Laor}}{1991}]{Laor1991}
{Laor} A.,  1991, \mn@doi [\apj] {10.1086/170257}, \href
  {https://ui.adsabs.harvard.edu/abs/1991ApJ...376...90L} {376, 90}

\bibitem[\protect\citeauthoryear{{Laplace, E.}, {G\"otberg, Y.}, {de Mink, S.
  E.}, {Justham, S.}  \& {Farmer, R.}}{{Laplace, E.}
  et~al.}{2020}]{Laplace2020}
{Laplace, E.} {G\"otberg, Y.} {de Mink, S. E.} {Justham, S.}  {Farmer, R.}
  2020, \mn@doi [A\&A] {10.1051/0004-6361/201937300}, 637, A6

\bibitem[\protect\citeauthoryear{Lazarus et~al.,}{Lazarus
  et~al.}{2013}]{Lazarus2013}
Lazarus P.,  et~al., 2013, \mn@doi [Monthly Notices of the Royal Astronomical
  Society] {10.1093/mnras/stt1996}, 437, 1485

\bibitem[\protect\citeauthoryear{{Mirabel, I. F.} \& {Rodrigues, I.}}{{Mirabel,
  I. F.} \& {Rodrigues, I.}}{2003}]{Mirabel2003}
{Mirabel, I. F.} {Rodrigues, I.} 2003, \mn@doi [A\&A]
  {10.1051/0004-6361:20021767}, 398, L25

\bibitem[\protect\citeauthoryear{Motta, Kajava, Sánchez-Fernández, Giustini
  \& Kuulkers}{Motta et~al.}{2017}]{Motta2017}
Motta S.~E.,  Kajava J. J.~E.,  Sánchez-Fernández C.,  Giustini M.,
  Kuulkers E.,  2017, \mn@doi [Monthly Notices of the Royal Astronomical
  Society] {10.1093/mnras/stx466}, 468, 981

\bibitem[\protect\citeauthoryear{{Parikh}, {Wijnands}  \&
  {Altamirano}}{{Parikh} et~al.}{2020}]{Parikh2020a}
{Parikh} A.~S.,  {Wijnands} R.,   {Altamirano} D.,  2020, The Astronomer's
  Telegram, \href {https://ui.adsabs.harvard.edu/abs/2020ATel13725....1P}
  {13725, 1}

\bibitem[\protect\citeauthoryear{Parker, Buisson, Tomsick, Fabian, Madsen,
  Walton  \& Fürst}{Parker et~al.}{2019}]{Parker2019}
Parker M.~L.,  Buisson D. J.~K.,  Tomsick J.~A.,  Fabian A.~C.,  Madsen K.~K.,
  Walton D.~J.,   Fürst F.,  2019, \mn@doi [Monthly Notices of the Royal
  Astronomical Society] {10.1093/mnras/stz045}, 484, 1202

\bibitem[\protect\citeauthoryear{{Parmar}, {White}, {Giommi}  \&
  {Gottwald}}{{Parmar} et~al.}{1986}]{Parmar1986}
{Parmar} A.~N.,  {White} N.~E.,  {Giommi} P.,   {Gottwald} M.,  1986, \mn@doi
  [\apj] {10.1086/164490}, \href
  {https://ui.adsabs.harvard.edu/abs/1986ApJ...308..199P} {308, 199}

\bibitem[\protect\citeauthoryear{{Parmar}, {Smale}, {Verbunt}  \&
  {Corbet}}{{Parmar} et~al.}{1991}]{Parmar1991}
{Parmar} A.~N.,  {Smale} A.~P.,  {Verbunt} F.,   {Corbet} R.~H.~D.,  1991,
  \mn@doi [\apj] {10.1086/169557}, \href
  {https://ui.adsabs.harvard.edu/abs/1991ApJ...366..253P} {366, 253}

\bibitem[\protect\citeauthoryear{Podsiadlowski}{Podsiadlowski}{1991}]{Podsiadlowski1991a}
Podsiadlowski P.,  1991, \mn@doi [Nature] {10.1038/350136a0}, 350, 136

\bibitem[\protect\citeauthoryear{Podsiadlowski}{Podsiadlowski}{2014}]{PPBinary}
Podsiadlowski P.,  2014, The evolution of binary systems.
Cambridge University Press, p. 45–88, \mn@doi{10.1017/CBO9781139343268.003}

\bibitem[\protect\citeauthoryear{Ponti, Fender, Begelman, Dunn, Neilsen  \&
  Coriat}{Ponti et~al.}{2012}]{Ponti2012}
Ponti G.,  Fender R.~P.,  Begelman M.~C.,  Dunn R. J.~H.,  Neilsen J.,   Coriat
  M.,  2012, \mn@doi [Monthly Notices of the Royal Astronomical Society:
  Letters] {10.1111/j.1745-3933.2012.01224.x}, 422, L11

\bibitem[\protect\citeauthoryear{Ponti, Muñoz-Darias  \& Fender}{Ponti
  et~al.}{2014}]{Ponti2014}
Ponti G.,  Muñoz-Darias T.,   Fender R.~P.,  2014, \mn@doi [Monthly Notices of
  the Royal Astronomical Society] {10.1093/mnras/stu1742}, 444, 1829

\bibitem[\protect\citeauthoryear{Postnov \& Yungelson}{Postnov \&
  Yungelson}{2014}]{Postnov2014}
Postnov K.~A.,  Yungelson L.~R.,  2014, \mn@doi [Living Reviews in Relativity]
  {10.12942/lrr-2014-3}, 17, 3

\bibitem[\protect\citeauthoryear{Puls, Vink  \& Najarro}{Puls
  et~al.}{2008}]{Puls2008}
Puls J.,  Vink J.~S.,   Najarro F.,  2008, \mn@doi [The Astronomy and
  Astrophysics Review] {10.1007/s00159-008-0015-8}, 16, 209

\bibitem[\protect\citeauthoryear{{Revnivtsev, M.}, {Gilfanov, M.}, {Churazov,
  E.}  \& {Sunyaev, R.}}{{Revnivtsev, M.} et~al.}{2002}]{Revnivtsev2002}
{Revnivtsev, M.} {Gilfanov, M.} {Churazov, E.}  {Sunyaev, R.} 2002, \mn@doi
  [A\&A] {10.1051/0004-6361:20020865}, 391, 1013

\bibitem[\protect\citeauthoryear{{Saikia}, {Russell}, {Baglio}, {Bramich}  \&
  {Lewis}}{{Saikia} et~al.}{2020}]{Saikia2020}
{Saikia} P.,  {Russell} D.~M.,  {Baglio} M.~C.,  {Bramich} D.~M.,   {Lewis} F.,
   2020, The Astronomer's Telegram, \href
  {https://ui.adsabs.harvard.edu/abs/2020ATel13719....1S} {13719, 1}

\bibitem[\protect\citeauthoryear{Steeghs \& Casares}{Steeghs \&
  Casares}{2002}]{Steeghs2002}
Steeghs D.,  Casares J.,  2002, \mn@doi [The Astrophysical Journal]
  {10.1086/339224}, 568, 273

\bibitem[\protect\citeauthoryear{Steiner, Lattimer  \& Brown}{Steiner
  et~al.}{2010}]{Steiner2010}
Steiner A.~W.,  Lattimer J.~M.,   Brown E.~F.,  2010, \mn@doi [The
  Astrophysical Journal] {10.1088/0004-637x/722/1/33}, 722, 33

\bibitem[\protect\citeauthoryear{{Stevens} \& {Uttley}}{{Stevens} \&
  {Uttley}}{2017}]{Stevens2017}
{Stevens} A.,  {Uttley} P.,  2017, in American Astronomical Society Meeting
  Abstracts \#229. p. 207.06

\bibitem[\protect\citeauthoryear{{Vasilopoulos}, {Bailyn}  \&
  {Milburn}}{{Vasilopoulos} et~al.}{2018}]{Vasilopoulos2018}
{Vasilopoulos} G.,  {Bailyn} C.,   {Milburn} J.,  2018, The Astronomer's
  Telegram, \href {https://ui.adsabs.harvard.edu/abs/2018ATel12164....1V}
  {12164, 1}

\bibitem[\protect\citeauthoryear{Verbunt}{Verbunt}{1993}]{Verbunt1993}
Verbunt F.,  1993, \mn@doi [Annual Review of Astronomy and Astrophysics]
  {10.1146/annurev.aa.31.090193.000521}, 31, 93

\bibitem[\protect\citeauthoryear{Walton et~al.,}{Walton
  et~al.}{2017}]{Walton2017}
Walton D.~J.,  et~al., 2017, \mn@doi [The Astrophysical Journal]
  {10.3847/1538-4357/aa67e8}, 839, 110

\bibitem[\protect\citeauthoryear{Wang et~al.,}{Wang et~al.}{2021}]{Wang2021}
Wang J.,  et~al., 2021, \mn@doi [The Astrophysical Journal Letters]
  {10.3847/2041-8213/abec79}, 910, L3

\bibitem[\protect\citeauthoryear{Wijnands \& van~der Klis}{Wijnands \& van~der
  Klis}{2000}]{Wijnands2000}
Wijnands R.,  van~der Klis M.,  2000, \mn@doi [The Astrophysical Journal]
  {10.1086/312439}, 528, L93

\bibitem[\protect\citeauthoryear{Wilms, Allen  \& McCray}{Wilms
  et~al.}{2000}]{Wilms2000}
Wilms J.,  Allen A.,   McCray R.,  2000, \mn@doi [The Astrophysical Journal]
  {10.1086/317016}, 542, 914

\bibitem[\protect\citeauthoryear{Wolff, Ray, Wood  \& Hertz}{Wolff
  et~al.}{2009}]{Wolff2009}
Wolff M.~T.,  Ray P.~S.,  Wood K.~S.,   Hertz P.~L.,  2009, \mn@doi [The
  Astrophysical Journal Supplement Series] {10.1088/0067-0049/183/1/156}, 183,
  156

\bibitem[\protect\citeauthoryear{{Yoon}, {Gr{\"a}fener}, {Vink}, {Kozyreva}  \&
  {Izzard}}{{Yoon} et~al.}{2012}]{Yoon2012}
{Yoon} S.~C.,  {Gr{\"a}fener} G.,  {Vink} J.~S.,  {Kozyreva} A.,   {Izzard}
  R.~G.,  2012, \mn@doi [\aap] {10.1051/0004-6361/201219790}, \href
  {https://ui.adsabs.harvard.edu/abs/2012A&A...544L..11Y} {544, L11}

\bibitem[\protect\citeauthoryear{Zhang, Makishima, Sakurai, Sasano  \&
  Ono}{Zhang et~al.}{2014}]{Zhang2013}
Zhang Z.,  Makishima K.,  Sakurai S.,  Sasano M.,   Ono K.,  2014, \mn@doi
  [Publications of the Astronomical Society of Japan] {10.1093/pasj/psu117}, 66

\bibitem[\protect\citeauthoryear{van~der Helm, Saladino, Portegies~Zwart  \&
  Pols}{van~der Helm et~al.}{2019}]{vanderHelm}
van~der Helm E.,  Saladino M.~I.,  Portegies~Zwart S.,   Pols O.,  2019,
  \mn@doi [A\&A] {10.1051/0004-6361/201732020}, 625, A85

\bibitem[\protect\citeauthoryear{van den Eijnden et~al.,}{van den Eijnden
  et~al.}{2020}]{VanDenEijnden2020}
van den Eijnden J.,  et~al., 2020, \mn@doi [Monthly Notices of the Royal
  Astronomical Society] {10.1093/mnras/staa1704}, 496, 4127

\bibitem[\protect\citeauthoryear{Özel \& Freire}{Özel \&
  Freire}{2016}]{Ozel2016}
Özel F.,  Freire P.,  2016, \mn@doi [Annual Review of Astronomy and
  Astrophysics] {10.1146/annurev-astro-081915-023322}, 54, 401

\makeatother
\end{thebibliography}

% Alternatively you could enter them by hand, like this:
% This method is tedious and prone to error if you have lots of references
% \begin{thebibliography}{99}
% \bibitem[\protect\citeauthoryear{Author}{2012}]{Author2012}
% Author A.~N., 2013, Journal of Improbable Astronomy, 1, 1
% \bibitem[\protect\citeauthoryear{Others}{2013}]{Others2013}
% Others S., 2012, Journal of Interesting Stuff, 17, 198
% \end{thebibliography}

%%%%%%%%%%%%%%%%%%%%%%%%%%%%%%%%%%%%%%%%%%%%%%%%%%

%%%%%%%%%%%%%%%%% APPENDICES %%%%%%%%%%%%%%%%%%%%%
\begin{appendices} 
\section{Markov Chain Monte Carlo}
\label{sx:ap1}
\begin{figure*}
\centering
\includegraphics[width=\textwidth]{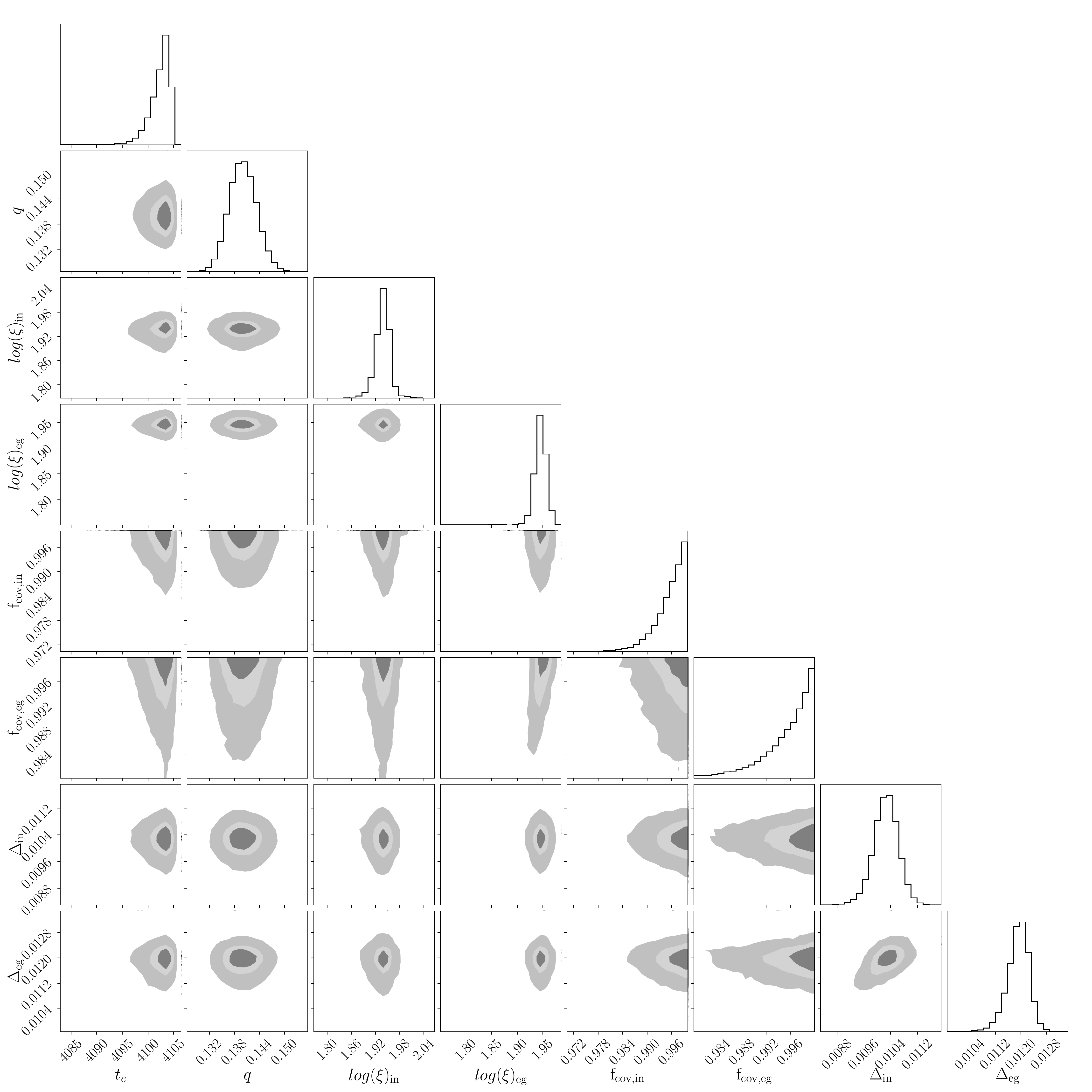}
\caption{Parameter distributions obtained by running a Markov Chain Monte Carlo (MCMC) simulation of the eclipse profile model assuming the Gaussian radial density profile. The MCMC is carried out within \textsc{xspec} and uses the Goodman-Weare algorithm. The chain has a length of 768000, 256 walkers and a burn-in period of 742912. For the 2D histograms, $1 \sigma$, $2 \sigma$ and $3 \sigma$ contours are respectively shaded in grey, silver and light grey. The 1D histograms are displayed with their y-axes in arbitrary units.
%Output distributions from the MCMC simulation of the eclipse profile model assuming the Gaussian radial density profile. The MCMC is run within \textsc{xspec} using the Goodman-Weare algorithm, a chain length of 768000, a burn-in period of 742912 and 256 walkers. The lines and shading, dark to light, on the 2D histograms represent $1 \sigma$, $2 \sigma$ and $3 \sigma$ contours respectively. The y-axes for the 1D histograms are in arbitrary units.
}
\label{fig:cornerGa}
\end{figure*}

\begin{figure*}
\centering
\includegraphics[width=\textwidth]{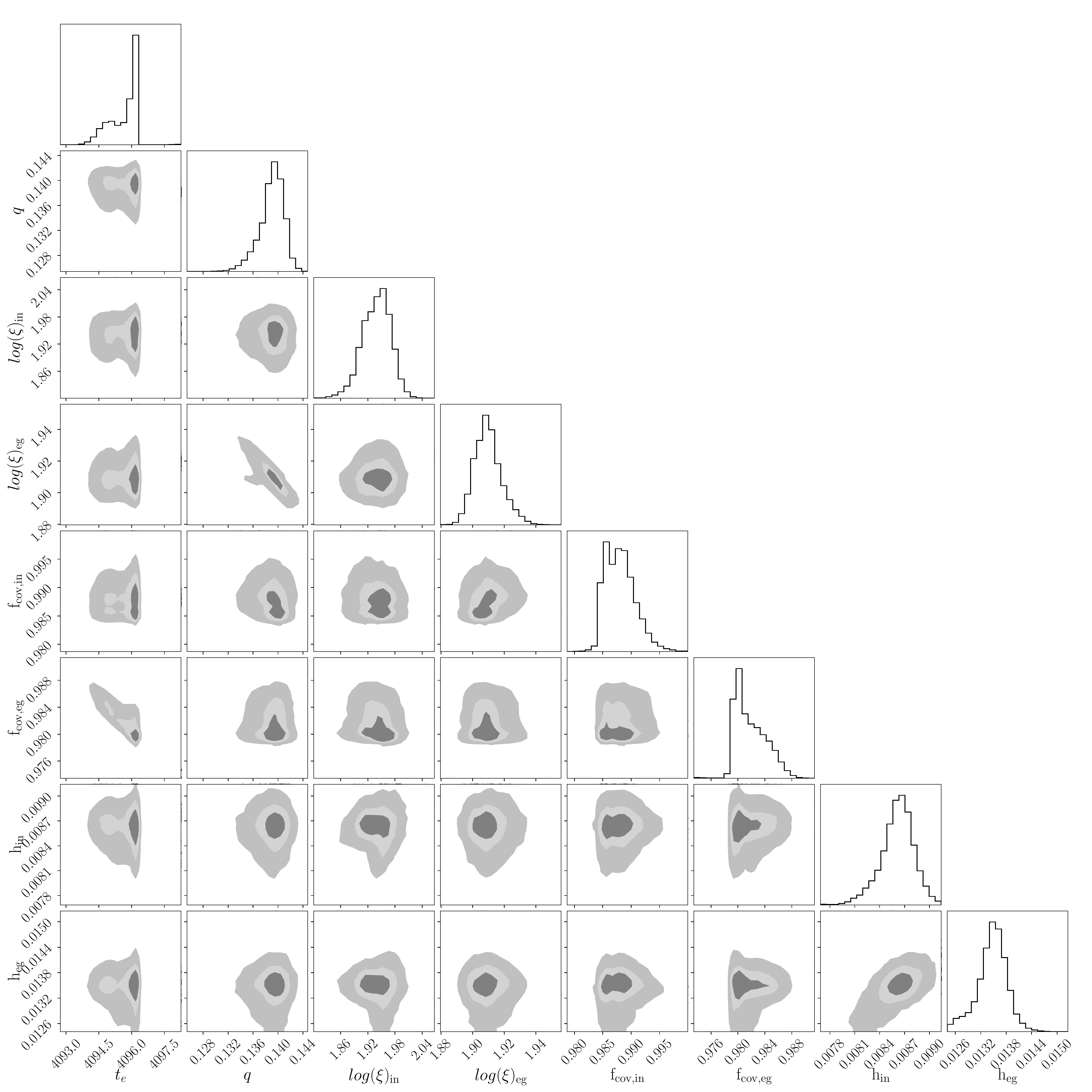}
\caption{Parameter distributions obtained by running a Markov Chain Monte Carlo (MCMC) simulation of the eclipse profile model assuming the exponential radial density profile. The MCMC is carried out within \textsc{xspec} and uses the Goodman-Weare algorithm. The chain has a length of 768000, 256 walkers and a burn-in period of 742912. For the 2D histograms, $1 \sigma$, $2 \sigma$ and $3 \sigma$ contours are respectively shaded in grey, silver and light grey. The 1D histograms are displayed with their y-axes in arbitrary units.
%Output distributions from the MCMC simulation of the eclipse profile model assuming the exponential radial density profile. The MCMC is run within \textsc{xspec} using the Goodman-Weare algorithm, a chain length of 768000, a burn-in period of 742912 and 256 walkers. The lines and shading, dark to light, on the 2D histograms represent $1 \sigma$, $2 \sigma$ and $3 \sigma$ contours respectively. The y-axes for the 1D histograms are in arbitrary units.
}
\label{fig:cornerEx}
\end{figure*}

To improve our understanding of the eclipse profile model parameter space, we run a Markov Chain Monte Carlo (MCMC) simulation within \textsc{xspec}. For each assumed density profile, we run the simulation with 256 walkers a chain length of 768000 and a burn-in period of 742912 steps using the Goodman-Weare algorithm. We start the chains from the best fitting parameters presented in Table \ref{tb:fitpars}. Figures \ref{fig:cornerGa} and \ref{fig:cornerEx} show the output distributions for each model parameter, for the eclipse profile model assuming the Gaussian and exponential radial density profiles respectively.

We do not find evidence for strong parameter correlations in the resulting distributions in either density profile model with one exception - the mass ratio, $q$ appears anti-correlated with egress ionisation $\log{\xi}_{\rm{eg}}$ when assuming the exponential density profile. In addition, there is some indication that the width of the material layer on the ingress side of the companion star is correlated with the material layer on the egress side of the companion star. This can be seen for both assumed density profiles. 

%The convergence of each chain is tested using the Geweke convergence measure which compares the mean of each parameter in two intervals of the chain, one shortly after the burn-in period and one towards the end of the chain. These correspond to the first $10 \%$ and the last $50 \%$ of the chain. For all chains, all parameters measured between $\pm 0.2$, indicating that convergence has been achieved.
The Geweke convergence measure was used to check that each of the MCMC simulations achieved convergence. This is carried out by comparing the mean of each parameter in the first $10 \%$ of the chain (i.e. shortly after the burn-in) and the last $50 \%$ of the chain. For both chains, we determined Geweke values in the range $\pm 0.2$, which suggests that convergence has been reached.

\end{appendices}

%%%%%%%%%%%%%%%%%%%%%%%%%%%%%%%%%%%%%%%%%%%%%%%%%%

% Don't change these lines
\bsp	% typesetting comment
\label{lastpage}
\end{document}